\documentclass[12pt]{article}
\usepackage{}
\usepackage{amsfonts}
\usepackage{bbm}
\usepackage{bm}
\usepackage{mathrsfs}
\usepackage{color}
\usepackage[left=2.30cm, right=2.30cm,top=2.0cm,bottom=2.0cm]{geometry}             
\usepackage{graphicx}
\usepackage{amssymb}
\usepackage{amsmath}
\usepackage{cite}
\usepackage{comment}
\usepackage{tikz}
\usetikzlibrary{matrix}

\usepackage[hyperindex=true,
          pdfstartview=FitH,
          bookmarksnumbered=true,
          bookmarksopen=true,
          citecolor=blue,
          linkcolor=blue,
          colorlinks=true,
          urlcolor=rossoCP3,
          pdfborder=001,
          unicode]{hyperref}

\usepackage{xcolor}  
\definecolor{mygreen}{RGB}{44,85,17}
\definecolor{myblue}{RGB}{34,31,217}
\definecolor{mybrown}{RGB}{194,164,113}
\definecolor{myred}{RGB}{255,66,56}
\definecolor{mypurple}{RGB}{200,36,176}

\newcommand{\mJ}{\mathcal{Y}}

\newcommand{\re}{\mathrm{e}}

\allowdisplaybreaks[4]
\definecolor{rossoCP3}{cmyk}{0,.88,.77,.40}

\begin{document}

\title{Linear response in a charged gas in curved spacetime \\
and covariant heat equation}
\author{Long Cui${}^1$, Xin Hao${}^{2,3}$, and Liu Zhao${}^1$
\vspace{1pt}\\
\small ${}^1$School of Physics, Nankai University, Tianjin 300071, China\\
\small ${}^2$School of Physics, Hebei Normal University, Shijiazhuang 050024, China\\
\small ${}^3$Shijiazhuang Key Laboratory of Astronomy and Space Science, Shijiazhuang 050024, China\\
\small {email}: \href{mailto:cuilong@mail.nankai.edu.cn}{cuilong@mail.nankai.edu.cn},\\
\small \href{mailto:xhao@hebtu.edu.cn}{xhao@hebtu.edu.cn}
\small and \href{mailto:lzhao@nankai.edu.cn}{lzhao@nankai.edu.cn}}
\date{ }
\maketitle

\begin{abstract}
  We consider the linear response of a near-equilibrium charged relativistic gas in
  the presence of electromagnetic and gravitational field in a generic stationary
  spacetime up to the second order of relaxation time
  and calculate the tensorial kinetic coefficients introduced by the presence
  of the strong electromagnetic and/or gravitational field.
  Using the covariant transfer equations thus developed, a covariant heat equation
  governing the relativistic heat conduction is derived,
  which, in Minkowski spacetime, reduces into a form which is remarkably
  similar to the well-known Cattaneo equation but with a different sign in front of
  the second-order time derivative term. We also perform a comparative analysis
  on the different behaviors of our heat equation and the Cattaneo equation in
  Minkowski spacetime. Furthermore, the effect of gravity
  on the heat conduction predicted by our heat equation
  is illustrated around Schwarzschild black hole, which makes a sharp contrast to
  the Minkowski case.

\end{abstract}

\section{Introduction}

In the nonrelativistic regime, Fick's law and Fourier's law are typically adopted
in describing diffusion and heat conduction, in which the particle flow $\vec j$
and the heat flow $\vec q$ are, respectively, considered to be
linear responses to the concentration and temperature gradients,
\begin{align*}
\vec j = -D \nabla n, \qquad \qquad
\vec q = -\kappa \nabla T.
\end{align*}
In the presence of long-range interactions such as electromagnetic and gravitational
interactions, the above gradients need to be replaced.
When incorporating electric fields, the particle flow
responds to the gradient of the electrochemical potential. Additionally,
in gravitational fields, the influence of temperature gradient will be
modulated by the Tolman-Ehrenfest effect \cite{Tolman:1930zza,Tolman:1930ona}, which
posits that gravity can offset temperature gradients to establish equilibrium.
Specifically, under Newtonian approximations, this relationship is described by
\begin{align*}
\nabla T = - T\, \nabla \phi.
\end{align*}
The quest to reconcile Fourier's law with the Tolman effect and relativistic effects has
motivated further investigations about relativistic thermodynamics\cite{Sandoval-Villalbazo:2012hkv}. However,
due to the lack of direct experimental verification in this field,
coupled with the challenges arising from nonequilibrium thermodynamics itself,
the study of gravitothermal effects is an ongoing endeavor \cite{Rovelli:2010mv,Lima:2019brf,Kim:2021kou,Majhi:2023stp,Xia:2023idh}.
A noteworthy progress in this area is the Luttinger theory \cite{luttinger1964theory} of thermal transport,
where gravitational field is introduced as a source term coupled to energy density.
The significance of the Luttinger theory lies in its ability to determine thermal transport coefficients through the application of the Kubo formula.
Furthermore, the linear transport equations provide additional insights,
\begin{equation}
\begin{split}
&\vec j_{n} = \bm L^{(1)} \left(\vec E - T\nabla \frac{\mu}{T}\right)
+ \bm L^{(2)} \left( { - \nabla \phi + T\nabla \frac{1}{T}}\right),   \\
&\vec j_{\epsilon} = \bm L^{(3)} \left(\vec E - T\nabla \frac{\mu}{T}\right)
+ \bm L^{(4)} \left({ - \nabla \phi + T\nabla \frac{1}{T}}\right),
\end{split}  \label{Luteq}
\end{equation}
where $ \vec j_{n} $ and $ \vec j_{\epsilon} $ denote the particle number flux and the energy flux, respectively.
The thermodynamic force in these equations is consistent with Tolman effect in the
Newtonian limit, Moreover, a crucial aspect is that the flux-force relation satisfies
the Onsager principle \cite{Onsager:1931jfa,Onsager:1931kxm}.
This is the beginning for gravity to appear in condensed-matter physics, and the seminal work
of Luttinger has now been generalized to the quantum level \cite{Chernodub:2021nff}, where various new transport effects
arise due to anomalies. In all these cases, only weak gravitational field is needed.

To extend the Luttinger theory to strong gravitational fields, relativistic kinetic theory
stands out as the most efficient and convenient approach \cite{juttner1911,Tauber:1961lbq,israel1963,
Rioseco:2016jwc,Mach:2021zqe,Sarbach2021}.
Kinetic theory is a common tool for modeling near-equilibrium fluid systems, allowing for systematic
calculations under the relaxation time approximation \cite{Hao:2021ifw,Liu:2022wpu,Hao:2023xgq}. While previous studies have primarily focused on neutral
fluids in weak gravitational fields at the first order of relaxation time, it becomes crucial to account for
higher-order relaxation time contributions in scenarios involving charged fluids under strong electromagnetic
fields or fluids interacting with strong gravitational fields. The inclusion of higher-order relaxation
time terms is essential to capture the complete linear response of the system.
The impact of external field strength manifests in the kinetic coefficients,
unveiling novel linear response phenomena such as transverse transport phenomena
and transport flows triggered by temperature variations.

Our previous work {\cite{Cui:2023bqg}} concentrated on the linear response at the first order of relaxation time,
serving as the foundation for this study. This paper delves into the linear response of
a charged fluid system in a generic stationary spacetime, extending the analysis
to the second order of relaxation time. For simplicity, we focus on a gaseous system
consisting of massive particles. Analytical computations within this framework are
comprehensive and manifestly covariant. Consequently, a covariant generalization
of the Luttinger theory is derived and the traditional Luttinger theory
should be reinstated in the weak gravitational fields limit.

Throughout this paper, we adopt the following conventions. The dimension of the
spacetime is taken to be $d+1$, with the metric signature $(-, +, \cdots, +)$.
Greek indices such as $\mu, \nu, \cdots$ take values in the range  $0, 1, \cdots, d$,
and latin indices $i, j, \cdots$ take values $1, 2, \cdots, d$. These indices
are, respectively, {\em coordinate} indices of the spacetime and of the space, while
hatted indices $\hat{a},\hat{b} = \hat{0}, \hat{1}, \cdots, \hat{d}$ and
$\hat{i}, \hat{j},\cdots = \hat{1}, \hat{2}, \cdots, \hat{d}$ refer to
{\em frame} indices. In addition, we use the physical quantity with
an overbar (such as $\bar{f},
\bar{T}, \bar{\mu}, \cdots$) to emphasize the values in detailed balance.

\section{Detailed balance for charged gas system}

In the framework of relativistic kinetic theory, the primary characteristics of
macroscopic systems are typically delineated by the energy-momentum tensor
$T^{\mu\nu}$, the particle current $N^\mu$, and the entropy current $S^\mu$.
These primary variables can be expressed as integrals involving
the one particle distribution function (1PDF) $f (x, p)$ over momentum space \cite{groot1980relativistic,cercignani2002},
\begin{equation}
  N^{\mu} = c \int \ensuremath{\boldsymbol{\varpi}} p^{\mu} f, \label{N,T}
  \qquad T^{\mu  \nu} = c \int \ensuremath{\boldsymbol{\varpi}} p^{\mu} p^{\nu} f,
\end{equation}
\begin{equation}
  S^{\mu} = - k_B c \int \ensuremath{\boldsymbol{\varpi}} p^{\mu} f \left[
  \log \left( \frac{h^d f}{f^{\ast}   \mathfrak{g}} \right) -
  \frac{\mathfrak{g} \log f^{\ast}}{\varsigma h^d f} \right], \label{S}
\end{equation}
where, as usual, $k_B$ is the Boltzmann constant, $h$ denotes the Planck constant,
$\mathfrak{g}$ stands for the degree of degeneracy, and $\varsigma=0, +1, -1$, respectively,
correspond to the case of nondegenerate, Bose-Einstein, and Fermi-Dirac statistics.
To streamline our formulas, we designated
$f^{\ast} = 1 + \varsigma \mathfrak{g}^{- 1} h^d f$ and
$\ensuremath{\boldsymbol{\varpi}} = \frac{\sqrt{g}}{| p_0 |}
(\mathrm{d} p)^d$ is the invariant volume element in the momentum space,
wherein $g=|\mathrm{det}(g_{\mu\nu})|$ with $g_{\mu\nu}$ being the
spacetime metric.

The 1PDF $f(x, p)$ holds a central position in kinetic theory. The evolution of the
1PDF is governed by the relativistic Boltzmann equation,
\begin{equation}
  \mathcal{L}_F f =\mathcal{C} (x, p),
\end{equation}
where the Liouville vector field $\mathcal{L}_F$ on the
tangent bundle of the spacetime \cite{Sarbach2012,Sarbach:2013hna} is given by
\begin{equation}
  \mathcal{L}_F = p^{\mu} \frac{\partial}{\partial x^{\mu}} + q
  F^{\mu}_{\enspace \nu} p^{\nu}  \frac{\partial}{\partial p^{\mu}} -
  \Gamma^{\mu}_{\alpha \beta} p^{\alpha} p^{\beta}  \frac{\partial}{\partial
  p^{\mu}}, \label{Liouville vector field}
\end{equation}
in which $q$ represents the charge of a particle and $F_{\mu \nu}$ is
the antisymmetric electromagnetic tensor, $\Gamma^{\mu}_{\alpha \beta}$ represents the
Christoffel connection associated with $g_{\mu\nu}$, and
the collision integral $\mathcal{C} (x, p)$ can be very complicated.
Under the assumption that two-body scatterings are dominating and that the
scatterings are microscopically reversible, the collision integral can be formulated as
\begin{equation}
  \mathcal{C} \left( {x, p_1}  \right) = \int
  \ensuremath{\boldsymbol{\varpi}}_2  \ensuremath{\boldsymbol{\varpi}}_3
  \ensuremath{\boldsymbol{\varpi}}_4  [W_x (p_3 + p_4 \leftrightarrow p_1 + p_2)  (f_3
  f_4 f_1^{\ast} f_2^{\ast} - f_1 f_2 f_3^{\ast} f_4^{\ast})],
  \label{collision term}
\end{equation}
where $W_x (p_3 + p_4 \leftrightarrow p_1 + p_2)$ represents the transition rate of
reversible two particles collision process $p_3 + p_4 \leftrightarrow p_1 + p_2$ that occurs
at the event $x$.

Without explicitly solving the relativistic Boltzmann equation,
we leverage the symmetry of the
collision integral to derive the entropy production in the form
\begin{equation}
  \nabla_{\mu} S^{\mu} = \frac{k_B c}{4} \int
  \ensuremath{\boldsymbol{\varpi}}_1  \ensuremath{\boldsymbol{\varpi}}_2
  \ensuremath{\boldsymbol{\varpi}}_3  \ensuremath{\boldsymbol{\varpi}}_4 W_x
  f_1 f_2 f_3^{\ast} f_4^{\ast} \left( \frac{f_3 f_4 f_1^{\ast}
  f_2^{\ast}}{f_1 f_2 f_3^{\ast} f_4^{\ast}} - 1 \right) \log \frac{f_3 f_4
  f_1^{\ast} f_2^{\ast}}{f_1 f_2 f_3^{\ast} f_4^{\ast}} .
  \label{entrprod}
\end{equation}
Clearly, the integrand in the above equation is always non-negative,
leading to $\nabla_{\mu} S^{\mu} \geqslant 0$, thereby illustrating the H-theorem.
Another important implication of the above equations is that, under the assumption
of microscopic reversibility, when entropy production is zero the collision integral
also vanishes. At this point, the system achieves detailed balance, and the associated 1PDF
$\bar f(x,p)$ is referred to as the detailed balance distribution, which takes the following form
\begin{equation}
  \bar{f} = \frac{\mathfrak{g}}{h^d}  \frac{1}{\re^{\bar{\alpha} -
  \bar{\mathcal{B}}_{\mu} p^{\mu}} - \varsigma}, \label{detailed balance}
\end{equation}
where $\bar{\alpha}$ is an additive scalar field arising from the collective
motion of particles and $\bar{\mathcal{B}}^{\mu}$ denotes a vector field.
Furthermore, given that $\bar{f}$ satisfies the Boltzmann equation with
vanishing collision integral, we have
\begin{equation}
  \mathcal{L}_F  \bar{f} = \left[ - p^{\mu} p^{\nu} \nabla_{\mu}
  \bar{\mathcal{B}}_{\nu} + p^{\mu}  \left( \nabla_{\mu} \bar{\alpha} - q
  F^{\nu}_{\enspace \mu}  \bar{\mathcal{B}}_{\nu} \right) \right]
  \frac{\partial \bar{f}}{\partial \bar{\alpha}} = 0.
\end{equation}
Considering the arbitrariness of the momenta, it follows that the first and
second order terms in the momenta should vanish separately. Consequently, we get,
for systems consisted of massive particles,
\begin{equation}
  \nabla_{\mu} \bar{\alpha} - q F^{\nu}_{\enspace \mu}   \bar{\mathcal{B}}_{\nu}=0,
  \qquad \nabla_{(\mu} \bar{\mathcal{B}}_{\nu)} = 0.
  \label{detailed balance condition}
\end{equation}
The second equality implies that $\bar{\mathcal{B}}^{\mu}$ is a Killing vector field.
In this work, our focus will be exclusively on systems comprising massive particles.
For systems consisting of massless particles, $\bar{\mathcal{B}}^{\mu}$ can be
conformal Killing.
Considering the nonrelativistic limit of detailed balance distribution,
the expression $- \bar{\mathcal B}_\mu p^\mu$ needs to be proportional to the energy
of the particle, which requires $\bar{\mathcal B}_\mu$ to be timelike.
Consequently, to uphold the existence of the detailed balance, $\bar{\mathcal{B}}^\mu$
must be a timelike Killing vector field, which in turn requires the spacetime
to be at least stationary. In this case, we can express
$\bar{\mathcal{B}}^{\mu}$ as $\bar{\mathcal{B}}^{\mu} = \bar{\beta}
\bar{U}^{\mu}$, where $\bar{U}^{\mu}  \bar{U}_{\mu} =
- c^2$. The normalized timelike vector field $\bar{U}^{\mu}$ represents the proper
velocity of the fluid in detailed balance. From the perspective of a comoving observer,
the local Euler relation Gibbs-Duhem relation for a charged system can be obtained,
\begin{equation}
  \bar{s} = k_B (\bar{\alpha}  \bar{n} + \bar{\beta}  \bar{\epsilon} +
  \bar{\beta}  \bar{P}),
\end{equation}
\begin{equation}
  - \bar{s} \mathrm{d} \left( \frac{1}{k_B  \bar{\beta}} \right) + \mathrm{d}
  \bar{P} + \bar{n} \mathrm{d} \left( \frac{\bar{\alpha}}{\bar{\beta}} \right)
  = 0,
\end{equation}
which imply that the thermodynamic correspondence holds for our charged system,
\begin{equation}
  \bar{\alpha} = - \frac{\bar{\mu}}{k_B  \bar{T}}, \quad \bar{\beta} =
  \frac{1}{k_B  \bar{T}}, \label{thermodynamic correspondence}
\end{equation}
where $\bar{\mu}$ is the chemical potential and $\bar{T}$ is the temperature
in detailed balance as measured by the comoving observer.

Relativistic kinetic theory is not limited to isolated single-component systems. It applies
to more general scenarios, such as multi-component systems in a thermal reservoir
or in the early Universe where reactions must be considered. In such cases,
the Boltzmann equation takes the form \cite{groot1980relativistic}
\begin{equation}
  \mathcal{L}_F f_A = \sum_B \mathcal{C}_{A B} (x, p_A),
\end{equation}
where the indices $A, B, \cdots$ designate different components of the fluid,
and the collision term becomes
\[ \mathcal{C}_{A B} \left( {x, p_A}  \right) = \frac{1}{2}  \sum_{C, D} \int
   \ensuremath{\boldsymbol{\varpi}}_B  \ensuremath{\boldsymbol{\varpi}}_C
   \ensuremath{\boldsymbol{\varpi}}_D  [W_x (p_C + p_D \mapsto p_A + p_B)
   (f_C f_D f_A^{\ast} f_B^{\ast} - f_A f_B f_C^{\ast} f_D^{\ast})] . \]
Through a comparable analysis, we can establish the same detailed balance conditions.

\section{Iterative solution of relativistic Boltzmann equation}

For nonequilibrium systems, there is no unique definition for the temperature,
chemical potential, and fluid velocity. When the system is slightly out of equilibrium,
the process of selecting a specific set of definitions for temperature,
chemical potential, and fluid velocity is known as a
choice of hydrodynamic frame \cite{Kovtun2012,Kovtun:2019hdm}. The choice of hydrodynamic frames does not alter
the primary variables $T^{\mu\nu}$ and $N^\mu$ but affects their hydrodynamic
expansion and thus highlights different aspects of interest.
Two commonly used hydrodynamic frames are Eckart frame and
Landau frame, in which the fluid velocity $U^\mu$ is taken respectively to be
parallel to the particle current and the unique timelike eigenvector of the
energy-momentum tensor.

Under the local equilibrium assumption for near-equilibrium systems, the zeroth-order 1PDF,
also known as the {\em local equilibrium distribution}, is typically assumed to take a form
analogous to the detailed balance distribution \eqref{detailed balance}, which
ensures the local thermodynamic relations.
\begin{equation}
  f^{(0)} = \frac{\mathfrak{g}}{h^d}  \frac{1}{\re^{\alpha -\mathcal{B}_{\mu}
  p^{\mu}} - \varsigma}, \label{local equilibrium}
\end{equation}
where $\mathcal{B}^{\mu}\equiv \beta U_\mu$ with $U^\mu$ normalized as
$U^\mu U_\mu = -c^2$, and the parameters $\alpha$ and $\beta$ are arbitrary.
Contrary to the case of detailed balance, the expressions
$\nabla_{\mu} \alpha - q F^{\sigma}_{\enspace \mu} \mathcal{B}_{\sigma}$
and $\nabla_{(\mu} \mathcal{B}_{\nu)}$ are now nonvanishing, and they characterize
the degree to which the system deviates from detailed balance.
Beginning from the next-to-leading order, different choices of hydrodynamic frame
impose distinct constraints, giving rise to different modifications to
$f^{(0)}$ at each order.

To facilitate hydrodynamic expansion, we can employ different collision models
in the context of kinetic theory. A commonly used collision model is based on
the relaxation time approximation, which simplifies the Boltzmann equation into
\begin{equation}
  \mathcal{L}_F f = - \frac{\varepsilon}{c^2 \tau} (f - f^{(0)}),
  \label{relt}
\end{equation}
where $\varepsilon = - U_{\mu} p^{\mu}$
and $\tau$ is known as the relaxation time, which represents
the timescale for the system to restore balance. It is important to emphasize
that writing down Eq.~\eqref{relt} does not imply a choice of a
concrete hydrodynamic frame. A hydrodynamic frame is fixed only when $U^\mu$ is
interpreted as the fluid velocity in an explicit hydrodynamic expansion of
$T^{\mu\nu}$ and $N^\mu$. For instance, when $U^\mu$ is identified with
the unique normalized timelike eigenvector of $T^{\mu\nu}$, we can say that
the Landau frame is chosen, and the collision model outlined in Eq.~\eqref{relt}
is known as the Anderson-Witting model \cite{anderson1974relativistic}. In this case, both the
particle current and the energy-momentum tensor satisfy conservation equations,
commonly employed in describing an isolated system.
However, in pursuit of a more comprehensive examination of transport phenomena,
we shall not confine ourselves to the Landau frame.

Irrespective to the choice of hydrodynamic frame, we can always iteratively solve
the Boltzmann equation \eqref{relt}
and categorize the iterations according to the power of the relaxation time.
Let $f^{(i)}$ be the $i$th-order modification to the
local equilibrium distribution $f^{(0)}$ and $f_{[n]}=\sum_{i=0}^n f^{(i)}$ be the
solution after the $n$th iteration. Then the iteration procedure can be achieved via
the equation
\begin{equation}
  \mathcal{L}_F f_{[n-1]} = - \frac{\varepsilon}{c^2 \tau} (f_{[n]} - f^{(0)}).
\end{equation}
Starting from $n=1$, we arrive at formal solution at generic order $n$,
\begin{equation}
  f_{[n]} = \sum_{i = 0}^n f^{(i)} = \sum_{i = 0}^n \left( -\frac{c^2
  \tau}{\varepsilon} \mathcal{L}_F \right)^i f^{(0)} .
  \label{iteration sequence}
\end{equation}

To analyze the transport phenomena at the linear response level, we need to
calculate the first few modification terms explicitly. To begin with,
at the order $\mathcal O(\tau)$, we have
\begin{align}
  f^{(1)} =  - \frac{c^2 \tau}{\varepsilon} \mathcal{L}_F f^{(0)}
  =  - \frac{c^2 \tau}{\varepsilon} \left[ p^{\mu}  \left( \nabla_{\mu} \alpha - q
  F^{\nu}_{\enspace \mu} \mathcal{B}_{\nu} \right) - p^{\mu} p^{\nu} \nabla_{\mu}
  \mathcal{B}_{\nu} \right]  \frac{\partial
  f^{(0)}}{\partial \alpha}, \label{f(1)}
\end{align}
in which the expressions $\nabla_{\mu} \alpha
- q F^{\sigma}_{\enspace \mu} \mathcal{B}_{\sigma}$
and $\nabla_{(\mu} \mathcal{B}_{\nu)}$ unequivocally signify the
departure from detailed balance.
Therefore, we designate them as thermodynamic forces,
\begin{align*}
\Psi_\mu = \nabla_{\mu} \alpha - q F^{\sigma}_{\enspace \mu} \mathcal{B}_{\sigma},
\qquad \Psi_{\mu\nu} = \nabla_{(\mu} \mathcal{B}_{\nu)}.
\end{align*}
It is evident that the entire first-order term in $\tau$ contributes
only to linear transport. However, linear transport is not contributed only from
the first-order term in $\tau$. In fact, linear transport contains contributions
from all orders in the relaxation time expansion of the 1PDF.
In this paper, we will focus on the linear response up to the second
order in relaxation time. Following Eq. \eqref{iteration sequence} and
neglecting some higher-order terms in the thermodynamic forces, we observe that
the terms that contribute to linear response at the order
$\mathcal O(\tau^2)$ can be divided into two parts,
\begin{align}
f^{(2)} = f^{(2)}_{\text{EM}} + f^{(2)}_{\text{G}}, \label{f(2)}
\end{align}
where
\begin{align}
  f^{(2)}_{\text{EM}}
  &\equiv \frac{\tau c^2 q}{\varepsilon^2} p^\sigma E_\sigma f^{(1)}
  + \left( \frac{c^2  \tau}{\varepsilon} \right)^2 q F^{\mu}_{\enspace \rho}
  \left( p^{\rho} \, \Psi_\mu - 2 p^{\rho} p^{\nu} \, \Psi_{\mu\nu} \right)
  \frac{\partial f^{(0)}}{\partial \alpha}, \\
  f^{(2)}_{\text{G}}
  &\equiv \frac{\tau}{\varepsilon} p^\sigma E^{(G)}_\sigma f^{(1)}
  + \left( \frac{c^2 \tau}{\varepsilon} \right)^2
  \left( p^{\sigma} p^{\mu} \nabla_{\sigma}
  \Psi_\mu - p^{\sigma} p^{\mu} p^{\nu} \nabla_{\sigma}
  \Psi_{\mu\nu}\right)
  \frac{\partial f^{(0)}}{\partial \alpha}.
  \label{f2gem}
\end{align}
The electric field $E_\mu$ and the gravitoelectric field $E^{(G)}_\mu$ are defined as
\begin{align}
E_{\mu} \equiv F_{\mu \sigma} U^{\sigma}, \qquad
E^{(G)}_{\mu} \equiv - U^{\rho} \nabla_{\rho} U_{\mu},
\end{align}
and, for later reference, we also introduce the magnetic field $B_{\mu\nu}$ and
the gravitomagnetic field $B^{(G)}_{\mu\nu}$ \footnote{
In more general situations, the vector field $U_\mu$ appearing in the definition
of the gravitoelectric field and the gravitomagnetic field can be replaced
by the proper velocity of some prescribed observer, indicating that the
gravitoelectromagnetic fields are actually observer dependent
and hence subject to Farady's effect. In this work, when $U_\mu$ is identified
as the proper velocity of the fluid element, we have actually implicitly
chosen the comoving observer. In such a setting,
for a steady fluid in the weak field limit, the metric and the fluid velocity
are given by
\[
\mathrm{d} s^2 = - c^2 \left( 1 + \frac{2 \phi}{c^2} \right) \mathrm{d} t^2
+ \frac{4}{c} A_i \mathrm{d} t \mathrm{d} x^i + \left( 1 - \frac{2
\phi}{c^2} \right) \delta_{i j} \mathrm{d} x^i \mathrm{d} x^j, \qquad
U^{\mu} = \left( 1 - \frac{\phi}{c^2} \right)  \left(
\frac{\partial}{\partial t} \right)^{\mu} .
\]
It can be verified that the gravitoelectromagnetic fields $E^{(G)}_{\mu}$ and
$B_{\mu \nu}^{(G)}$ reduce to their traditional definitions $E^{(G)}_i = -
\partial_i \phi, B_{i j}^{(G)} = 2 \partial_{[i} A_{j]}$.
} as
\begin{align}
B_{\mu \nu} \equiv F_{\rho \sigma} \Delta^{\rho}_{\enspace \mu}
\Delta^{\sigma}_{\enspace \nu}, \qquad
B_{\mu \nu}^{(G)} \equiv \Delta^{\sigma}_{\enspace \mu}
  \Delta^{\rho}_{\enspace \nu} \nabla_{[\sigma} U_{\rho]},
\end{align}
where $\Delta^{\mu\nu} = g^{\mu\nu}+c^{-2} U^\mu U^\nu$ denotes
the projection tensor.
The decomposition of $f^{(2)}$ is motivated by the distinct origins of transport phenomena,
one emanating from the electromagnetic field and the other from the gravitational field.
Later on, we shall see that the terms involving covariant derivatives of
$\Psi_\mu$ and $\Psi_{\mu\nu}$ in $ f^{(2)}_{\text{G}}$ can be decomposed into
terms that contain linear, derivative and product expressions in the irreducible
thermodynamic forces. Since we are concentrating on the linear response regime, the terms
containing derivative and product expressions in the irreducible
thermodynamic forces will be disregarded.

\section{Linear response}

This section is devoted to the calculation of the particle current and the
energy-momentum tensor up to $\mathcal O(\tau^2)$. At the leading order,
the form of $N^\mu$ and $T^{\mu\nu}$ is similar to that for a perfect fluid
in detailed balance,
\begin{align}
  N^{\mu(0)} =  \mJ_{d - 1, 1}
  U^{\mu}, \qquad
  T^{\mu\nu(0)} = c^{-2} \mJ_{d - 1, 2}
  U^{\mu} U^{\nu} + d^{-1} \mJ_{d + 1, 0}
  \Delta^{\mu \nu}. \label{N0T0}
\end{align}
It is evident that up to $\mathcal O (\tau^0)$ the system is fully characterized
by scalar densities, namely, the particle number density, energy density,
and pressure
\begin{align}
n^{(0)} = \mJ_{d - 1, 1}, \qquad \epsilon^{(0)} = \mJ_{d - 1, 2},
\qquad P = d^{-1} \mJ_{d + 1, 0}  \label{n0e0P},
\end{align}
signifying no transport flux presence.
These scalars are functions in $(\alpha,\beta)$, and they can all be defined
as integrals of the type
\begin{align}
  \mJ_{n, \ell} (\alpha, \beta) \equiv \frac{\mathfrak{g}}{\lambda_C^d}
  \mathcal{A}_{d - 1} \left(mc^2\right)^{n+\ell-d}
  \int_0^{\infty} \frac{\sinh^n \vartheta
  \cosh^\ell \vartheta}{\re^{\alpha + \beta mc^2 \cosh \vartheta} - \varsigma} \mathrm{d}
  \vartheta, \label{specialfunction2}
\end{align}
in which $\mathcal{A}_{d - 1}$ represents the area of the $(d - 1)$-dimensional
unit sphere and $\lambda_C=h/(mc)$ is the Compton wavelength with $m$ being the mass
of the particle. It can be verified that the function
$\mJ_{n, \ell}(\alpha,\beta)$ obeys the following relation:
\begin{align}
\frac{\partial}{\partial \alpha} \mJ_{n, \ell} (\alpha,\beta)
= \frac{\partial}{\partial \beta}\mJ_{n, \ell - 1} (\alpha,\beta). \label{partail_y relation}
\end{align}

By substituting the first-order Eq.~\eqref{f(1)} and the second-order
Eq.~\eqref{f(2)} modifications of the 1PDF into the expressions for the
particle current and energy-momentum tensor \eqref{N,T} while disregarding
nonlinear terms in the thermodynamic forces, we derive a set
of linear response equations,
\begin{align}
\left( \begin{matrix}
		\delta N^\mu \\
		\delta T^{\mu\nu} \\
\end{matrix} \right)
=
\left( \begin{matrix}
	L^{\mu\sigma} & L^{\mu\sigma\rho} \\
	L^{\mu\nu\sigma} & L^{\mu\nu\sigma\rho} \\
\end{matrix} \right)
\left( \begin{matrix}
		\Psi_\sigma \\
		\Psi_{\sigma\rho} \\
\end{matrix} \right), \label{covarFF}
\end{align}
where $\delta N^\mu$ and $\delta T^{\mu\nu}$ denote the deviations of the particle current
and energy-momentum tensor at the linear order in the thermodynamic forces $\Psi_\mu$ and
$\Psi_{\mu\nu}$. The tensorial linear response equation \eqref{covarFF} is among
the main result of this work. When the thermodynamic forces are weak,
linear response effects dominate, with the kinetic coefficients
characterizing the nonequilibrium properties. These kinetic coefficients can be
calculated through kinetic theory. For instance, at order $\mathcal{O}(\tau)$, we have
\begin{align*}
\left( \begin{matrix}
	L^{\mu\sigma} \\
    L^{\mu\sigma\rho} \\
	L^{\mu\nu\sigma\rho} \\
\end{matrix} \right)^{(1)}
= c \, \frac{\partial}{\partial \alpha}
\int\boldsymbol{\varpi} f^{(0)} \left(\frac{c^2\tau}{\varepsilon}\right)
\left( \begin{matrix}
	p^\mu p^\sigma \\
    p^\mu p^\sigma p^\rho \\
	p^\mu p^\nu p^\sigma p^\rho \\
\end{matrix} \right),
\end{align*}
and results for the kinetic coefficients at order $\mathcal{O}(\tau^2)$ are organized in Appendix A.
Equation \eqref{covarFF} elegantly describes the flux-force relation in a covariant manner,
wherein the kinetic coefficients clearly satisfy the Onsager reciprocal relation.
However, in order to gain a phenomenological understanding of Eq.~\eqref{covarFF},
it is necessary to decompose it into irreducible parts.
On the left-hand side, this amounts to
\begin{align}
\{ \delta N^\mu, \delta T^{\mu\nu} \} \to
\{(\delta n, \, \delta \epsilon, \, \Pi), \,
(j^\mu, \, q^\mu), \, \Pi^{\mu\nu}\},
\end{align}
with
\begin{align}
  \delta N^{\mu} & =  \delta n \, U^{\mu} + j^{\mu}, \nonumber\\
  \delta T^{\mu \nu} & =  \frac{1}{c^2} \delta \epsilon \, U^{\mu} U^{\nu} + \frac{1}{c^2}
  q^{\mu} U^{\nu} + \frac{1}{c^2} q^{\nu} U^{\mu} + \Pi \Delta^{\mu \nu}
  + \Pi^{\mu \nu}. \label{NT}
\end{align}
As a result, we obtain three scalar responses: the perturbation in particle number density $\delta n$,
the perturbation in energy density $\delta \epsilon$, and the dynamic pressure $\Pi$,
along with two vector responses (the particle flux $j^\mu$ and the energy flux $q^\mu$)
and one tensor response (the deviatoric stress tensor $\Pi^{\mu\nu}$).
Similarly, the thermodynamic forces can also be
decomposed into three irreducible types,
\begin{align}
\{\Psi_\mu, \Psi_{\mu\nu}\} \to
\{(\dot{\alpha}, \, \dot{\beta}, \,\psi), \,
(\mathcal D_\mu \alpha, \, \mathcal D_\mu \beta),
\, \psi_{\mu\nu}\},
\end{align}
in which the scalar thermodynamic forces are
\begin{align*}
\dot{\alpha} = U^{\mu} \nabla_{\mu} \alpha, \qquad
\dot{\beta} = U^{\mu} \nabla_{\mu} \beta, \qquad
\psi = - \beta \nabla_{\mu} U^{\mu},
\end{align*}
the vectorial thermodynamic forces are
\begin{align*}
\mathcal D_\mu \alpha = \left(\nabla_\nu \alpha
-qF^\sigma_{\enspace \nu} \mathcal B_\sigma\right) \Delta_{\enspace\mu}^\nu,
\quad \mathcal D_\mu \beta =
\left(\nabla_{\nu} \beta - \frac{\beta}{c^2} U^{\rho}
  \nabla_{\rho} U_{\nu} \right) \Delta_{\enspace\mu}^\nu,
\end{align*}
and tensorial thermodynamic force is
\begin{align*}
\psi_{\mu\nu} = - \beta \left( \Delta_{\mu}^{\enspace \rho} \Delta_{\nu}^{\enspace \sigma}
\nabla_{(\rho} U_{\sigma)} - \dfrac{1}{d} \nabla_{\rho} U^{\rho} \Delta_{\mu
\nu} \right).
\end{align*}

Now, it is the right place to work out the linear response equations for the
irreducible components. First,
due to the fulfillment of the Onsager relation by the overall covariant flux-force relation,
it is reasonable to anticipate that each irreducible deviation, when responding to thermodynamic
force of the same type, will satisfy the Onsager reciprocal principle. Such response equations can be
classified into three categories:

(i) The scalar-scalar response (S-S) is
\begin{align}\small
&\left( \begin{matrix}
		\delta n \\
		\delta \epsilon \\
        \Pi
\end{matrix} \right)
= - \tau \frac{\partial}{\partial \alpha}
\left( \begin{matrix}
	\mJ_{d-1, 1} & \mJ_{d-1, 2} & d^{-1} \mJ_{d+1,0} \\
	\mJ_{d-1, 2} & \mJ_{d-1, 3} & d^{-1} \mJ_{d+1,1} \\
    d^{-1} \mJ_{d+1,0} & d^{-1} \mJ_{d+1,1} & d^{-2} \mJ_{d+3,-1} \\
\end{matrix} \right)
\left( \begin{matrix}
		\dot{\alpha} \\
		\dot{\beta} \\
        \psi
\end{matrix} \right), \label{S-S}
\end{align}
\indent (ii) The vector-vector response (V-V) is
\begin{align}\small
\left( \begin{matrix}
		j^\mu \\
		q^\mu \\
\end{matrix} \right)
= - \tau c^2 \frac{1}{d}
\left[\left(\Delta^{\mu\nu}+ \tau B_{(G)}^{\mu\nu}\right)\frac{\partial}{\partial \beta}
+ \tau c^2 q B^{\mu\nu}\frac{\partial}{\partial \alpha}\right]
\left( \begin{matrix}
	\mJ_{d+1, -2} & \mJ_{d+1, -1} \\
	\mJ_{d+1, -1} & \mJ_{d+1, 0} \\
\end{matrix} \right)
\left( \begin{matrix}
		\mathcal D_\nu \alpha \\
		\mathcal D_\nu \beta \\
\end{matrix} \right), \label{V-V}
\end{align}
\indent (iii) The tensor-tensor response (T-T) is
\begin{align}\small
\Pi^{\mu\nu}
= - \tau \frac{2}{(d+2)d}
\left[\left(\Delta^{\rho\mu}\Delta^{\nu\sigma}
+2\tau \Delta^{\rho(\mu} B_{(G)}^{\nu)\sigma} \right)
\frac{\partial}{\partial \beta}
+2\tau c^2 q \, \Delta^{\rho(\mu} B^{\nu)\sigma}
\frac{\partial}{\partial \alpha}\right] \mJ_{d+3,-2}
\, \psi_{\sigma \rho}
\end{align}
In addition to the aforementioned linear response equations, perturbations can also respond
to thermodynamic forces of different types: \\
\indent (i) The scalar-vector response (S-V) is
\begin{align}\small
&\left( \begin{matrix}
		\delta n \\
		\delta \epsilon \\
        \Pi
\end{matrix} \right)
=
\tau^2 \left(E_{(G)}^{\mu} \frac{\partial}{\partial \beta}
+ c^2 q E^{\mu} \frac{\partial}{\partial \alpha}\right)
\left( \begin{matrix}
	\mJ_{d-1, 0} - d^{-1} \mJ_{d+1,-2} & \mJ_{d-1, 1}  \\
	\mJ_{d-1, 1} - d^{-1} \mJ_{d+1,-1} & \mJ_{d-1, 2} \\
    d^{-1} \mJ_{d+1, -1} - d^{-2} \mJ_{d+3,-3} & d^{-1}\mJ_{d+1,0} \\
\end{matrix} \right)
\left( \begin{matrix}
		\mathcal{D}_{\mu} \alpha \\
		\mathcal{D}_{\mu} \beta \\
\end{matrix} \right).  \label{S-V}
\end{align}
\indent (ii) The vector-scalar response (V-S) is
\begin{align}\small
\left( \begin{matrix}
		j^\mu \\
		q^\mu \\
\end{matrix} \right)
= \tau^2 \frac{1}{d}
\left(E_{(G)}^{\mu}\frac{\partial}{\partial \beta}
+ c^2 q E^{\mu}\frac{\partial}{\partial \alpha}\right)
\left( \begin{matrix}
	0 & \mJ_{d+1, -1} & 2d^{-1} \mJ_{d+1, -1} - d^{-1} \mJ_{d+3, -3}  \\
	0 & \mJ_{d+1, 0} & 2d^{-1} \mJ_{d+1, 0} - d^{-1} \mJ_{d+3, -2} \\
\end{matrix} \right)
\left( \begin{matrix}
		\dot \alpha \\
		\dot \beta \\
        \psi \\
\end{matrix} \right), \label{V-S}
\end{align}
\indent (iii) The vector-tensor response (V-T) is
\begin{align}\small
\left( \begin{matrix}
		j^\mu \\
		q^\mu \\
\end{matrix} \right)
= \tau^2 \frac{2}{d}
\left(E^{(G)}_{\rho}\frac{\partial}{\partial \beta}
+ c^2 q E_{\rho}\frac{\partial}{\partial \alpha}\right)
\left( \begin{matrix}
	\mJ_{d+1, -1} - (d+2)^{-1} \mJ_{d+3, -3}  \\
	\mJ_{d+1, 0} - (d+2)^{-1} \mJ_{d+3, -2}  \\
\end{matrix} \right)
\psi^{\rho\mu}, \label{V-T}
\end{align}
\indent (iv) The tensor-vector response (T-V) is
\begin{align}\small
\Pi^{\mu\nu}
= - \tau^2 \frac{2}{(d+2)d}
\left( \Delta^{\rho (\mu} \Delta^{\nu) \sigma} -
  \frac{1}{d} \Delta^{\rho \sigma} \Delta^{\mu \nu} \right)
\left(E^{(G)}_{\rho}\frac{\partial}{\partial \beta}
+ c^2 q E_\rho \frac{\partial}{\partial \alpha}\right) \mJ_{d+3,-3}
\, \mathcal D_\sigma \alpha. \label{T-V}
\end{align}
Please be reminded that the derivatives $\partial/\partial \alpha,
\partial/\partial \beta$ in the above  equations act only on the function
$\mJ_{n, \ell} (\alpha, \beta)$.

By examining the irreducible linear response equations provided above,
it can be noted that the V-V response can be interpreted as the covariant
Luttinger equation, which in the Newtonian limit
reduces to its conventional form \eqref{Luteq}. The T-T response determines
the shear viscosity, while the S-S response, upon selecting a hydrodynamic frame,
can be used to calculate the bulk viscosity.
For the cross-type responses, it is evident that all kinetic coefficients are
proportional to $ E^{\mu} $ and $ E^{\mu}_{(G)} $, which are the sources of
inhomogeneities in chemical potential and temperature
for the equilibrium system. Therefore, it is reasonable to infer that cross-type
responses arise from these inhomogeneities and disappear when both $ E^{\mu} $ and $ E^{\mu}_{(G)} $ vanish.

\section{Covariant heat equation}\label{Applications}

In the nonrelativistic context, heat conduction is conventionally described by the Fourier
equation, which is a parabolic partial differential equation,
\begin{align}
\frac{\partial T}{\partial t}
= \frac{\kappa}{c_V} \nabla^2 T,
\end{align}
where $c_V$ is the heat capacity per unit volume.
The Fourier equation has faced criticism due to its nature as a parabolic partial differential equation,
which appears to imply an infinite speed of heat conduction, thus violating the principle of causality.
However, it is crucial to acknowledge that the Fourier equation is solely applicable in the realm of
linear transport. Any violation of causality resulting from large thermodynamic forces falls beyond
the scope of the Fourier equation and does not undermine its validity.
Anyway, it is necessary to generalize heat conduction to its relativistic counterpart.
In the subsequent discussion, we aim to comprehend the heat conduction by utilizing linear
response equations derived from relativistic kinetic theory.

In standard textbooks, the heat equation can be derived through energy conservation.
Taking a similar approach, we initiate our discussion by considering the
divergence-free condition of energy-momentum tensor
to the first order of relaxation time
\begin{align}
\nabla_\mu \left(T^{\mu\nu(0)} + \delta T^{\mu\nu(1)}\right) = 0,  \label{Conserv-O1}
\end{align}
where $T^{\mu\nu(0)}$ represents the local equilibrium energy-momentum tensor \eqref{N0T0}
and $\delta T^{\mu \nu(1)}$ denotes the first-order perturbation in \eqref{NT}.
Contracting Eq. \eqref{Conserv-O1} with $U_\nu$, we obtain
\begin{align}
U^{\mu} \nabla_{\mu} \epsilon^{(0)} + (\epsilon^{(0)} + P) \nabla_{\mu} U^{\mu}
+ U^{\mu} \nabla_{\mu} \delta \epsilon^{(1)} + \nabla_{\mu} q^{\mu(1)} +
\frac{1}{c^2} q^{\nu(1)} U^{\mu} \nabla_{\mu} U_{\nu} =0 \label{gghe}
\end{align}
which, according to the constitutive relations \eqref{n0e0P}, \eqref{S-S}, and \eqref{V-V},
is an equation involving variables  $\left(\alpha ,\beta, \dot{\alpha}, \dot{\beta},
\mathcal D_\mu \alpha, \mathcal D_\mu \beta \right)$. To further simplify Eq.~\eqref{gghe}
toward the heat equation, it is both reasonable and necessary to consider the following
two assumptions:

\noindent (i) The particle number density remains invariant $U^\mu \nabla_\mu n^{(0)} = 0$,
and the expansion of the fluid system is negligible $\nabla_\mu U^\mu = 0$.
Under these conditions, $\dot \alpha$ and $\dot \beta$ satisfy
\begin{align}
\frac{\partial \mathcal{Y}_{d - 1, 1}}{\partial \alpha}\dot{\alpha} + \frac{\partial \mathcal{Y}_{d - 1, 1}}{\partial
\beta} \dot{\beta} = 0,
\end{align}
allowing the heat capacity per unit volume to be characterized as
\begin{align}
U^{\mu} \nabla_{\mu} \epsilon^{(0)} = - \frac{1}{k_B T^2}
\left[ - \frac{\left(\frac{\partial \mathcal{Y}_{d - 1, 1}}{\partial
\beta}\right)^2}{\frac{\partial \mathcal{Y}_{d - 1, 0}}{\partial \beta}}
+ \frac{\partial \mathcal{Y}_{d - 1, 2}}{\partial \beta} \right] U^\mu \nabla_\mu T \equiv c_V \dot T.
\label{c_V}
\end{align}
\noindent ({ii}) The absence of particle flow, $ j^\mu = 0$,
correlates $\mathcal D_\mu \alpha$ and $\mathcal D_\mu \beta$ as
\begin{align}
\frac{\partial \mathcal{Y}_{d + 1, -1}}{\partial \alpha}\mathcal D_\mu \alpha
+ \frac{\partial \mathcal{Y}_{d + 1, -1}}{\partial\beta} \mathcal D_\mu \beta = 0
\label{heatCD}
\end{align}
thereby defining the heat conductivity
\begin{align}
q_\mu = \frac{\tau c^2}{d} \frac{1}{k_B T^2}
\left[ - \frac{\left(\frac{\partial \mathcal{Y}_{d + 1, -1}}{\partial
\beta}\right)^2}{\frac{\partial \mathcal{Y}_{d + 1, -2}}{\partial \beta}}
+ \frac{\partial \mathcal{Y}_{d + 1, 0}}{\partial \beta} \right] \mathcal D_\mu T
\equiv - \kappa \, \mathcal D_\mu T.
\end{align}
Incorporating the S-S response \eqref{S-S} and
V-V response \eqref{V-V}, along with Eqs.~\eqref{c_V} and \eqref{heatCD}, we finally
obtain the general covariant heat equation
\begin{align}
c_V \dot T - \tau c_V \ddot T = \kappa \, {\mathcal D}^2 T +
\frac{T}{c^2} \frac{\partial \kappa}{\partial T}
{E^\mu_{(g)}} \mathcal D_\mu T , \label{CovHE}
\end{align}
where
\begin{align*}
\dot T = U^\mu \nabla_\mu T, \qquad
\mathcal D_\mu T = \Delta_{\enspace\mu}^\nu
\left(\nabla_{\nu} + \frac{1}{c^2} U^{\rho} \nabla_{\rho} U_{\nu} \right) T \\
\ddot T = U^\mu \nabla_\mu \dot T,   \qquad
\mathcal D^2 T = \left(\nabla_{\mu} + \frac{1}{c^2} U^{\rho} \nabla_{\rho} U_{\mu} \right)
\mathcal D^\mu T~.
\end{align*}
The covariant heat equation \eqref{CovHE} demonstrates that gravitational effect is manifested
in both the alteration of spatial derivative operators and the additional term concerning
the gravitoelectric field.  For comparison with the Cattaneo equation,
we consider an inertial and static fluid in Minkowski spacetime,
simplifying Eq.~\eqref{CovHE} into
\begin{align}
c_V \dot T - \tau c_V \ddot T = \kappa \nabla^2 T, \label{Ell}
\end{align}
which exhibits a remarkable similarity to the Cattaneo equation
\begin{align}
c_V \dot T + \tau c_V \ddot T = \kappa \nabla^2 T
\label{Catta}
\end{align}
which has been known for over seventy years \cite{Cattaneo1948,Vernotte1958,Lopez-Monsalvo:2010oeo,Kim:2022qlc}.
The only difference of our heat equation
\eqref{Ell} from the Cattaneo equation \eqref{Catta} lies in a different sign
in front of the second-order time derivative term. This sign difference makes a sharp
contrast regarding the mathematical behaviors of the two equations. While the
Cattaneo equation is a hyperbolic partial differential equation, our heat equation
\eqref{Ell} is elliptic.
Notably, the second-order time derivative term in the Cattaneo equation
results from introducing a retardation effect in the flux-force relation,
thus surpassing the realm of linear response. Conversely, in deriving Eq.
\eqref{Ell}, we maintained ourselves in the linear response regime.
The presence of the second-order time derivative term
originates from considering $\delta \epsilon^{(1)}$, which can be
explicitly expressed through the S-S response.
The advantage of preserving the structure of linear response is obvious:
it allows for the precise control of thermodynamic forces to maintain
a timelike energy flow, thereby ensuring the principle of relativity.
It is also worth it to remember that the parameter $\tau$ in
Eqs.~\eqref{Ell} and \eqref{Catta} has different origins. The parameter $\tau$
in the Cattaneo equation is introduced phenomenologically as
a retardation time, while in our Eq. \eqref{Ell}, $\tau$ is clearly the
relaxation time for the relativistic gas.

\begin{figure}[htb!]
	\begin{center}
		\includegraphics[height=.3\textheight]{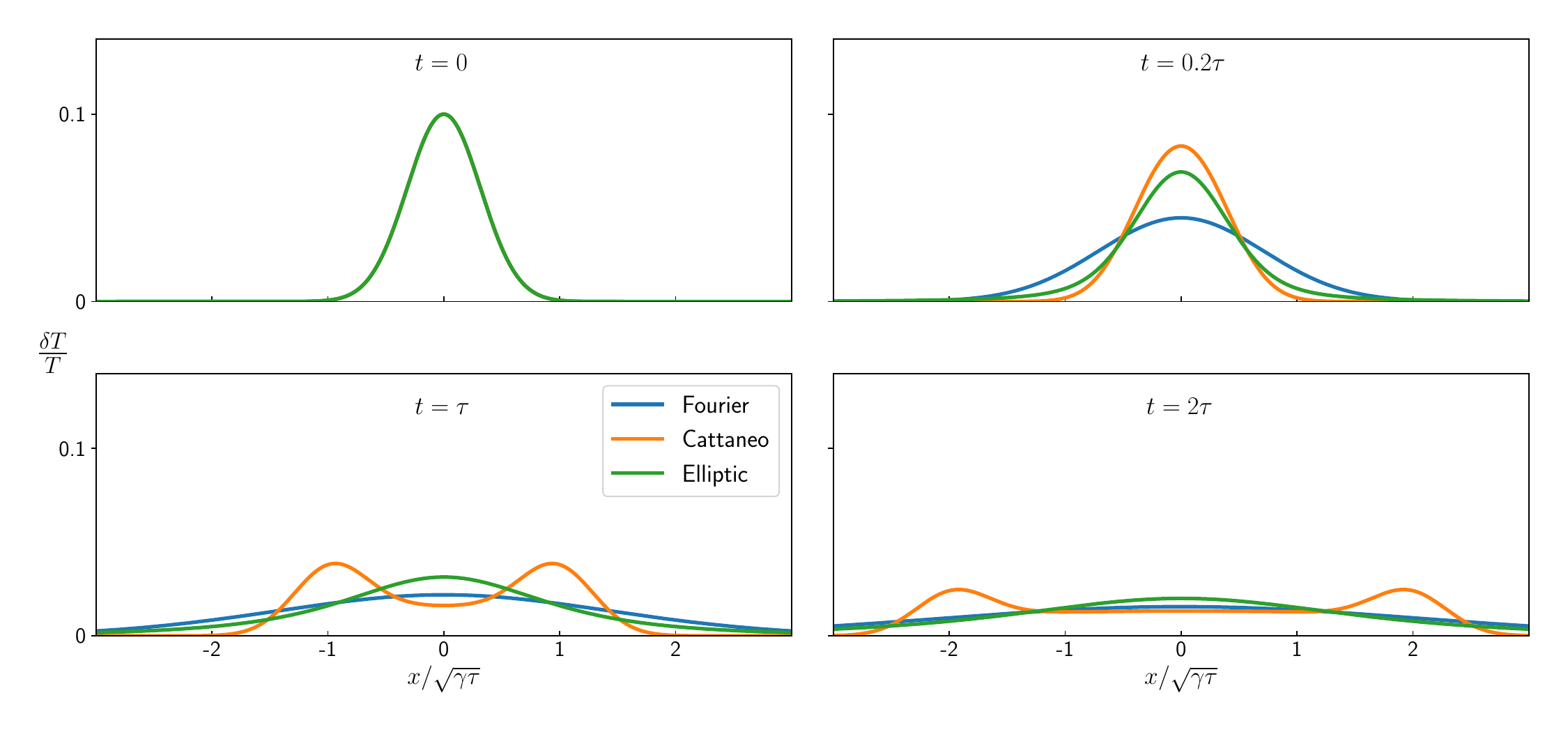}	
		
		\caption{Time evolution of a Gaussian-type temperature perturbation.}\label{PHEg}
	\end{center}
\end{figure}

To comprehend the implications of different heat equations,
we examine the solution of the equation
\begin{align}
\dot T + {\vartheta}\, \tau \, \ddot T - \gamma \, \nabla^2 T = 0, \label{EHE}
\end{align}
where $\gamma= \kappa/c_V$ is the thermal diffusivity, and $\vartheta = 0, +1, -1$
correspond to the
Fourier equation, the Cattaneo equation, and the elliptic heat equation, respectively.
When considering the perturbation with a typical wavelength $\lambda$, the
Fourier equation reveals that the temperature fluctuation decays exponentially with time:
\begin{align*}
\delta T_{\,\text{Fourier}} \sim \exp\left(-\frac{4\pi^2\gamma}{\lambda^2}t\right).
\end{align*}
In the case of the Cattaneo equation and the elliptic heat equation,
the evolution of temperature perturbations with long wavelengths
is consistent with the Fourier equation.
However, for perturbations with short wavelengths, the evolution described
by the three equations differs. The elliptic heat equation presents an evolution of
$\delta T$ similar to that prescribed
by the Fourier equation, featuring an exponential decay over time.
Specifically, in the short wavelength limit,
\begin{align*}
\delta T_{\,\text{Elliptic}} \sim
\exp\left(-\dfrac{2\pi}{\lambda}\sqrt{\dfrac{\gamma}{\tau}}\,t\right).
\end{align*}
Conversely, the Cattaneo equation reveals significant difference for
perturbations with wavelengths smaller than $4\pi\sqrt{\gamma\tau}$, wherein
$\delta T$ demonstrate oscillatory behavior,
\begin{align*}
\delta T_{\,\text{Cattaneo}} \sim
\exp\left({-\frac{t}{2\tau}}\right)
\cos\left(\dfrac{2\pi}{\lambda}\sqrt{\dfrac{\gamma}{\tau}}\,t\right).
\end{align*}
Such oscillation indicates that, upon reaching thermal equilibrium,
the evolution continues, allowing heat to be transferred from cooler to warmer
regions, which seemingly contradicts the principles of thermodynamics.
To provide an intuitive illustration, we conduct a numerical examination
of a Gaussian-type temperature perturbation and depict the result graphically
in Fig.\ref{PHEg}. In these plots, $T$ refers to the equilibrium
temperature of the gaseous system, and $\delta T$ is the local departure from the
equilibrium temperature.
As expected, noticeable differences in temperature evolution
are observed among the three types of heat equations.

\begin{figure}[!h]
	\begin{center}
		\includegraphics[height=0.55\textheight]{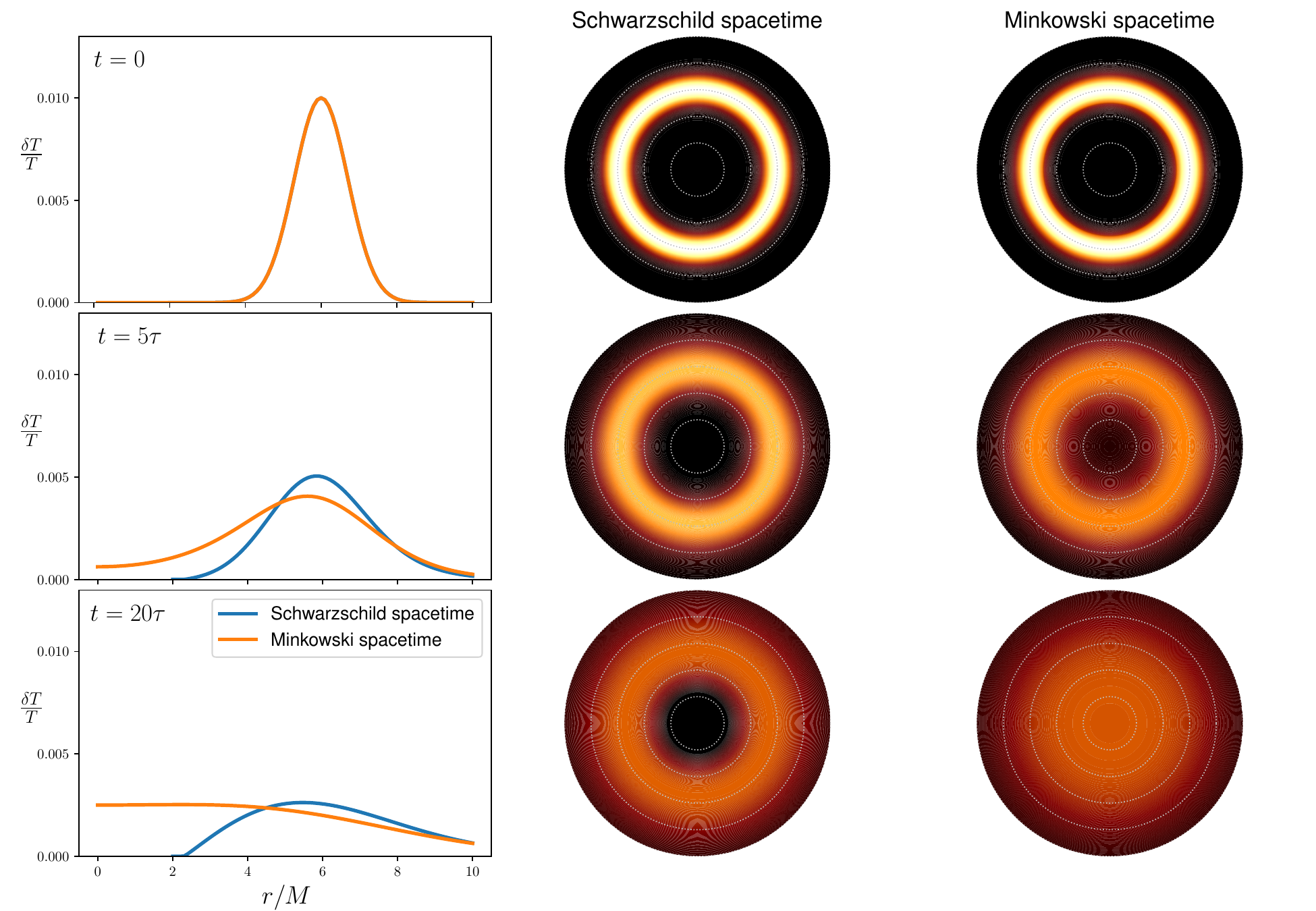}	
		
		\caption{Time evolution of a radial Gaussian-type temperature perturbation.}
    \label{RadEvolCP}
	\end{center}
\end{figure}

To illustrate the impact of gravity on heat conduction, we also make a comparison
between the evolution of a spherical temperature perturbation in
Minkowski spacetime and around the Schwarzschild black hole. For simplicity,
we employ a radial Gaussian distribution peaked at $r=6M$ as the
initial temperature profile, where $M$ denotes the mass of the
Schwarzschild black hole. Please be reminded that the equilibrium temperature $T$
and the local temperature departure $\delta T$ in the case of Schwarzschild
spacetime are both evaluated at
asymptotic infinity by the inclusion of a red shift factor.
After solving the covariant heat equation
\eqref{CovHE} numerically, we observe different temperature profiles
for the two different spacetimes
after the same period of heat transfer, as depicted in Fig.\ref{RadEvolCP}.
It is evident that thermal diffusion around black hole is influenced by the
gravitational redshift effect. Specifically, in terms of coordinate
time, the thermal diffusion slows down in the near horizon region.

\section{Conclusions}

We studied the linear response of a charged gaseous system in curved spacetime
subjected to an external electromagnetic field up to the second order of
relaxation time. Our findings extend previous results by
including linear response induced by both the electromagnetic
field and the gravitoelectromagnetic field.
The distinctive features of this study are summarized as follows:
\begin{itemize}
\item[(i)] The kinetic theory calculations yield a succinct tensorial flux-force equation \eqref{covarFF}.
  All phenomenological linear response equations can be derived as irreducible decompositions of
  this equation, including the covariant form of the renowned Luttinger theory.

\item[(ii)] In the presence of an electric field or a gravitoelectric field, the tensorial flux-force equation
  predicts novel cross-type responses \eqref{S-V}, \eqref{V-S},  \eqref{V-T}, and \eqref{T-V}.

\item[(iii)] A covariant heat equation \eqref{CovHE} is established from the perspective of kinetic theory, facilitating the
  theoretical investigation of thermal  diffusion in curved spacetime.

\item[(iv)] In Minkowski spacetime, the covariant heat equation simplifies to an elliptic-type equation \eqref{EHE}.
  Compared to the Cattaneo equation, the elliptic heat equation prevents temperature
  oscillations during thermal diffusion.
\end{itemize}

It is noteworthy that in the derivation of the tensorial
flux-force equation \eqref{covarFF} no particular
hydrodynamic frame is initially selected. Consequently,
when applying Eq. \eqref{covarFF},
we are free to choose the hydrodynamic frame according to the physical scenarios.
For example, when considering conservation laws, direct calculations indicate that
\[
\nabla_\mu \left(N^{\mu(0)} + N^{\mu(1)}\right)
= -\frac{1}{\tau}\left(\delta n^{(1)}+ \delta n^{(2)}\right),
\]
\[
\nabla_\mu \left(T^{\mu\nu(0)} + T^{\mu\nu(1)}\right) = -\frac{1}{c^2\tau}\left[
\left(\delta \epsilon^{(1)}+ \delta \epsilon^{(2)}\right)U^\nu
+ \left(q^{\nu(1)} + q^{\nu(2)}\right)\right]
+ q F^{\nu}{}_\mu\left(N^{\mu(0)} + N^{\mu(1)}\right).
\]
Then, for neutral systems, if the conservation equation is required up to
the first order of relaxation time, the corresponding constraints are
\[
\delta n^{(1)} + \delta n^{(2)} = 0, \qquad
\delta \epsilon^{(1)} +\delta \epsilon^{(2)} = 0, \qquad
q^{\mu(1)} + q^{\mu(2)} = 0,
\]
which can be regarded as the constraint conditions for the
Landau frame up to the second order of relaxation time.

Much has been left to do following the line of this work. The first question
is whether the tensorial flux-force equation is still valid
when a higher-order relaxation time effect is taken into account.
A comparison between the results at the $\mathcal O(\tau)$ and $\mathcal O(\tau^2)$
suggests that the higher-order effects may introduce nontrivial correction terms,
which motivates us to formulate linear response equations encompassing
higher-orders relaxation time.
Another direction centers on the application of the covariant heat equation.
The example provided in this study concerning spherically symmetric temperature
perturbations in Schwarzschild spacetime is oversimplified. Future work should
address more realistic scenarios by examining nonspherically symmetric initial
temperature perturbations and more general spacetime backgrounds, which may reveal
novel features of gravity-induced lateral heat transport.

\section*{Acknowledgement}

This work is supported by the National Natural Science
Foundation of China under Grant No. 12275138 and the Hebei NSF
under Grant No. A2021205037.

\section*{Conflict of interests declaration}

The authors declare no known conflict of interests.

\section*{Data availability}

No data were created or analyzed in this study.


\providecommand{\href}[2]{#2}\begingroup\raggedright\endgroup

\section*{Appendix A:~~~Kinetic coefficients}

To calculate the transport coefficients induced by gravity,
we need to extract the linear response from Eq. \eqref{f2gem}.
First, notice that for any tensor $\Phi_{\mu}, \Phi_{\mu \nu}, \Phi_{\mu \nu \rho}$
the following decomposition can be obtained by inserting
$\delta^{\alpha}_{\enspace \mu} = \Delta^{\alpha}_{\enspace \mu}
- c^{-2} U^{\alpha} U_{\mu}$:
\begin{align*}
  \Phi_{\mu}  = &~ \left( \Delta^{\alpha}_{\enspace \mu} - \frac{1}{c^2}
  U^{\alpha} U_{\mu} \right) \Phi_{\alpha}\\
   = &~ \Delta^{\alpha}_{\enspace \mu} \Phi_{\alpha} - \frac{1}{c^2} U_{\mu}
  U^{\alpha} \Phi_{\alpha},\\
  \Phi_{\mu \nu}  = &~ \left( \Delta^{\alpha}_{\enspace \mu} - \frac{1}{c^2}
  U^{\alpha} U_{\mu} \right) \left( \Delta^{\beta}_{\enspace \nu} -
  \frac{1}{c^2} U^{\beta} U_{\nu} \right) \Phi_{\alpha \beta}\\
   = &~ \Delta^{\alpha}_{\enspace \mu} \Delta^{\beta}_{\enspace \nu}
  \Phi_{\alpha \beta} - \frac{1}{c^2} U_{\mu} \Delta^{\beta}_{\enspace \nu}
  U^{\alpha} \Phi_{\alpha \beta} - \frac{1}{c^2} U_{\nu}
  \Delta^{\alpha}_{\enspace \mu} U^{\beta} \Phi_{\alpha \beta} + \frac{1}{c^4}
  U_{\mu} U_{\nu} U^{\alpha} U^{\beta} \Phi_{\alpha \beta},\\
  \Phi_{\mu \nu \rho}  = &~ \left( \Delta^{\alpha}_{\enspace \mu} -
  \frac{1}{c^2} U^{\alpha} U_{\mu} \right) \left( \Delta^{\beta}_{\enspace
  \nu} - \frac{1}{c^2} U^{\beta} U_{\nu} \right) \left(
  \Delta^{\gamma}_{\enspace \rho} - \frac{1}{c^2} U^{\gamma} U_{\rho} \right)
  \Phi_{\alpha \beta \gamma}\\
   = & - \frac{1}{c^6} U_{\mu} U_{\nu} U_{\rho} U^{\alpha} U^{\beta}
  U^{\gamma} \Phi_{\alpha \beta \gamma} + \Delta^{\alpha}_{\enspace \mu}
  \Delta^{\beta}_{\enspace \nu} \Delta^{\gamma}_{\enspace \rho} \Phi_{\alpha
  \beta \gamma}\\
    & - \frac{1}{c^2} U_{\mu} U^{\alpha} \Delta^{\beta}_{\enspace \nu}
  \Delta^{\gamma}_{\enspace \rho} \Phi_{\alpha \beta \gamma} - \frac{1}{c^2}
  U_{\nu} \Delta^{\alpha}_{\enspace \mu} U^{\beta} \Delta^{\gamma}_{\enspace
  \rho} \Phi_{\alpha \beta \gamma} - \frac{1}{c^2} U_{\rho}
  \Delta^{\alpha}_{\enspace \mu} \Delta^{\beta}_{\enspace \nu} U^{\gamma}
  \Phi_{\alpha \beta \gamma}\\
    & + \frac{1}{c^4} U_{\nu} U_{\rho} \Delta^{\alpha}_{\enspace \mu}
  U^{\beta} U^{\gamma} \Phi_{\alpha \beta \gamma} + \frac{1}{c^4} U_{\mu}
  U_{\rho} U^{\alpha} \Delta^{\beta}_{\enspace \nu} U^{\gamma} \Phi_{\alpha
  \beta \gamma} + \frac{1}{c^4} U_{\mu} U_{\nu} U^{\alpha} U^{\beta}
  \Delta^{\gamma}_{\enspace \rho} \Phi_{\alpha \beta \gamma} .
\end{align*}
Applying the above decomposition, we can decompose thermodynamic forces $\Psi_{\mu}, \Psi_{\mu \nu}$ into scalar, vectorial, and tensorial parts,
\begin{align}
  \Psi_{\lambda}  = & - \frac{1}{c^2}  \dot{\alpha} U_{\lambda}
  +\mathcal{D}_{\lambda} \alpha,\label{Psi1}\\
  \Psi_{\rho \lambda}  = & - \frac{1}{c^2}  \dot{\beta} U_{\rho} U_{\lambda}
  +\mathcal{D}_{(\rho} \beta U_{\lambda)} - \psi_{\rho \lambda} - \frac{1}{d}
  \psi \Delta_{\rho \lambda}.\label{Psi2}
\end{align}
And we also need to use the following important decomposition:
\begin{align*}
  \nabla_{\mu} U_{\nu}  = &~ \Delta^{\rho}_{\enspace \mu}
  \Delta^{\sigma}_{\enspace \nu} \nabla_{(\rho} U_{\sigma)} +
  \Delta^{\rho}_{\enspace \mu} \Delta^{\sigma}_{\enspace \nu} \nabla_{[\rho}
  U_{\sigma]} - \frac{1}{c^2} U_{\mu} U^{\rho} \nabla_{\rho} U_{\nu} \\
   = & - \frac{1}{\beta} \psi_{\mu \nu} - \frac{1}{d}  \frac{1}{\beta} \psi
  \Delta_{\mu \nu} + B_{\mu \nu}^{(G)} + \frac{1}{c^2} U_{\mu} E^{(G)}_{\nu}.
\end{align*}

By using Eqs.~\eqref{Psi1} and \eqref{Psi2}, we can decompose
$\nabla_{\sigma} \Psi_{\lambda}$ and $\nabla_{\sigma} \Psi_{\rho \lambda}$
in Eq.~\eqref{f2gem} into linear response terms, derivative terms of
thermodynamic forces, and product terms of thermodynamic forces,
\begin{align*}
  \nabla_{\sigma} \Psi_{\lambda}  = & - \frac{1}{c^2} \nabla_{\sigma}
  \dot{\alpha} U_{\lambda} + \nabla_{\sigma} \mathcal{D}_{\lambda} \alpha -
  \frac{1}{c^2}  \dot{\alpha} \left( - \frac{1}{\beta} \psi_{\sigma \lambda} -
  \frac{1}{d}  \frac{1}{\beta} \psi \Delta_{\sigma \lambda} + B_{\sigma
  \lambda}^{(G)} + \frac{1}{c^2} U_{\sigma} E^{(G)}_{\lambda} \right),\\
  \nabla_{\sigma} \Psi_{\rho \lambda}  = & - \frac{1}{c^2} \nabla_{\sigma}
  \dot{\beta} U_{\rho} U_{\lambda} - \frac{2}{c^2}  \dot{\beta}  \left( -
  \frac{1}{\beta} \psi_{\sigma (\rho} U_{\lambda)} - \frac{1}{d}
  \frac{1}{\beta} \psi \Delta_{\sigma (\rho} U_{\lambda)} + B_{\sigma
  (\rho}^{(G)} U_{\lambda)} + \frac{1}{c^2} U_{\sigma} E^{(G)}_{(\rho}
  U_{\lambda)} \right)\\
    & + \nabla_{\sigma} \mathcal{D}_{(\rho} \beta U_{\lambda)} -
  \frac{1}{\beta} \psi_{\sigma (\lambda} \mathcal{D}_{\rho)} \beta -
  \frac{1}{d}  \frac{1}{\beta} \psi \Delta_{\sigma (\lambda}
  \mathcal{D}_{\rho)} \beta + B_{\sigma (\lambda}^{(G)} \mathcal{D}_{\rho)}
  \beta + \frac{1}{c^2} U_{\sigma} E^{(G)}_{(\lambda} \mathcal{D}_{\rho)}
  \beta\\
    & - \nabla_{\sigma} \psi_{\rho \lambda} - \frac{1}{d} \nabla_{\sigma}
  \psi \Delta_{\rho \lambda} - \frac{1}{d} \psi \frac{2}{c^2} \left( -
  \frac{1}{\beta} \psi_{\sigma (\rho} U_{\lambda)} - \frac{1}{d}
  \frac{1}{\beta} \psi \Delta_{\sigma (\rho} U_{\lambda)} + B_{\sigma
  (\rho}^{(G)} U_{\lambda)} + \frac{1}{c^2} U_{\sigma} E^{(G)}_{(\rho}
  U_{\lambda)} \right),
\end{align*}
where $\nabla_{\sigma} \mathcal{D}_{\lambda} \alpha$,
$\nabla_{\sigma} \mathcal{D}_{\alpha} \beta$, and
$\nabla_{\sigma} \psi_{\rho \lambda}$ can be further decomposed as
\begin{align*}
  \nabla_{\sigma} \mathcal{D}_{\lambda} \alpha  = &~ \Delta^{\alpha}_{\enspace
  \sigma} \Delta^{\beta}_{\enspace \lambda} \nabla_{\alpha}
  \mathcal{D}_{\beta} \alpha - \frac{1}{c^2} U_{\sigma}
  \Delta^{\beta}_{\enspace \lambda} U^{\alpha} \nabla_{\alpha}
  \mathcal{D}_{\beta} \alpha + \frac{1}{c^2} U_{\lambda} \Delta^{\beta \alpha}
  \left( B^{(G)}_{\sigma \alpha} + \frac{1}{c^2} U_{\sigma} E^{(G)}_{\alpha}
  \right) \mathcal{D}_{\beta} \alpha\\
    & + \frac{1}{c^2} U_{\lambda} \mathcal{D}_{\beta} \alpha \Delta^{\beta
  \gamma} \Delta^{\alpha}_{\enspace \sigma} \nabla_{(\alpha} U_{\gamma)},\\
  \nabla_{\sigma} \mathcal{D}_{\alpha} \beta  = &~ \Delta^{\gamma}_{\enspace
  \sigma} \Delta^{\beta}_{\enspace \alpha} \nabla_{\gamma} \mathcal{D}_{\beta}
  \beta - \frac{1}{c^2} U_{\sigma} \Delta^{\beta}_{\enspace \alpha} U^{\gamma}
  \nabla_{\gamma} \mathcal{D}_{\beta} \beta + \frac{1}{c^2} U_{\alpha}
  \Delta^{\beta \gamma} \left( B^{(G)}_{\sigma \gamma} + \frac{1}{c^2}
  U_{\sigma} E^{(G)}_{\gamma} \right) \mathcal{D}_{\beta} \beta\\
    & + \frac{1}{c^2} U_{\alpha} \mathcal{D}_{\beta} \beta \Delta^{\beta
  \gamma} \Delta^{\delta}_{\enspace \sigma} \nabla_{(\gamma} U_{\delta)},\\
  \nabla_{\sigma} \psi_{\rho \lambda}  = &~ \Delta^{\alpha}_{\enspace \sigma}
  \Delta^{\beta}_{\enspace \rho} \Delta^{\gamma}_{\enspace \lambda}
  \nabla_{\alpha} \psi_{\beta \gamma} - \frac{1}{c^2} U_{\sigma} U^{\alpha}
  \Delta^{\beta}_{\enspace \rho} \Delta^{\gamma}_{\enspace \lambda}
  \nabla_{\alpha} \psi_{\beta \gamma}\\
    & + \frac{2}{c^2} \psi_{\beta \gamma} \Delta^{\eta \beta}
  \Delta^{\gamma}_{\enspace (\lambda} U_{\rho)} \left( - \frac{1}{\beta}
  \psi_{\sigma \eta} - \frac{1}{d}  \frac{1}{\beta} \psi \Delta_{\sigma \eta}
  + B_{\sigma \eta}^{(G)} + \frac{1}{c^2} U_{\sigma} E^{(G)}_{\eta} \right) .
\end{align*}
For ease of expression, we introduce the notation $\text{linear} [\cdots]$ to denote
the part of linear response terms contained in the square brackets.
Using this notation, we extract the linear response component in
$\nabla_{\sigma} \Psi_{\lambda}$ and $\nabla_{\sigma} \Psi_{\rho \lambda}$ as follows:
\begin{align}
  \text{linear} [\nabla_{\sigma} \Psi_{\lambda}]  = & - \frac{1}{c^2}
  \dot{\alpha} \left( B_{\sigma \lambda}^{(G)} + \frac{1}{c^2} U_{\sigma}
  E^{(G)}_{\lambda} \right) + \frac{1}{c^2} U_{\lambda} \Delta^{\beta \alpha}
  \left( B^{(G)}_{\sigma \alpha} + \frac{1}{c^2} U_{\sigma} E^{(G)}_{\alpha}
  \right) \mathcal{D}_{\beta} \alpha, \label{linear part1}\\
  \text{linear} [\nabla_{\sigma} \Psi_{\rho \lambda}]  = & - \frac{2}{c^2}
  \dot{\beta}  \left( B_{\sigma (\rho}^{(G)} U_{\lambda)} + \frac{1}{c^2}
  U_{\sigma} E^{(G)}_{(\rho} U_{\lambda)} \right) - \frac{1}{d} \psi
  \frac{2}{c^2} \left( B_{\sigma (\rho}^{(G)} U_{\lambda)} + \frac{1}{c^2}
  U_{\sigma} E^{(G)}_{(\rho} U_{\lambda)} \right) \nonumber\\
    & + F_{\sigma (\lambda}^{(G)} \mathcal{D}_{\rho)} \beta + \frac{1}{c^2}
  \Delta^{\beta \alpha} \left( B^{(G)}_{\sigma \alpha} + \frac{1}{c^2}
  U_{\sigma} E^{(G)}_{\alpha} \right) \mathcal{D}_{\beta} \beta \, U_{\rho}
  U_{\lambda} \nonumber\\
    & - \frac{2}{c^2} \psi_{\beta \gamma} \Delta^{\eta \beta}
  \Delta^{\gamma}_{\enspace (\lambda} U_{\rho)} \left( B_{\sigma \eta}^{(G)} +
  \frac{1}{c^2} U_{\sigma} E^{(G)}_{\eta} \right) \label{linear part2}.
\end{align}

Once the linear response has been decomposed, the next step is to perform the integration.
We work in orthonormal basis $\{(e_{\hat{a}})^{\mu} \}$ obeying
$\eta_{\hat{a} \hat{b}} = g_{\mu \nu}
(e_{\hat{a}})^{\mu}  (e_{\hat{b}})^{\nu}$. Without loss of
generality, we require that $U^{\mu} = c (e_{\hat{0}})^{\mu}$, which implies that the
induced metric can be expressed as $\Delta^{\mu \nu} = \delta^{\hat{i}
\hat{j}}  (e_{\hat{i}})^{\mu}  (e_{\hat{j}})^{\nu}$.

For massive particles, the momentum $p^{\hat{a}} = p^{\mu}
(e^{\hat{a}})_{\mu}$ can be parametrized by mass shell conditions
\begin{equation}
  p^{\hat{a}} = m c (\cosh \vartheta, n^{\hat{i}} \sinh \vartheta), \label{p
  parameter}
\end{equation}
where $n^{\hat{i}} \in S^{d - 1}$ is a spacelike unit vector. Then, the
momentum space volume element can be represented as
\begin{equation}
  \ensuremath{\boldsymbol{\varpi}} = \frac{(\mathrm{d} p)^d}{| p_{\hat{0}} |}
  = \frac{| \ensuremath{\boldsymbol{p}} |^{d - 1} \mathrm{d} |
  \ensuremath{\boldsymbol{p}} | \mathrm{d} \Omega_{d - 1}}{p^{\hat{0}}} = (m c
  \sinh \vartheta)^{d - 1} \mathrm{d} \vartheta \mathrm{d} \Omega_{d - 1},
\end{equation}
where $\mathrm{d} \Omega_{d - 1}$ is the volume element of the $(d
- 1)$-dimensional unit sphere $S^{d - 1}$. In this way, the
integration in the momentum space can be decomposed into integration over
$\vartheta \in (0, \infty)$ and integration over the unit sphere $S^{d - 1}$.

Here, we list some useful integration formulas,
\begin{align}
  \int \mathrm{d} \Omega_{d - 1} & =  \mathcal{A}_{d - 1}, \\
  \int n^{\hat{i}} n^{\hat{j}} \mathrm{d} \Omega_{d - 1} & =  \frac{1}{d}
  \mathcal{A}_{d - 1} \delta^{\hat{i}  \hat{j}}, \\
  \int n^{\hat{i}} n^{\hat{j}} n^{\hat{k}} n^{\hat{l}} \mathrm{d} \Omega_{d -
  1} & =  \frac{3}{(d + 2) d} \mathcal{A}_{d - 1} \delta^{(\hat{i}  \hat{j}}
  \delta^{\hat{k}  \hat{l} )},
\end{align}
\begin{equation}
  \int n^{\hat{i}} \mathrm{d} \Omega_{d - 1} = \int n^{\hat{i}} n^{\hat{j}}
  n^{\hat{k}} \mathrm{d} \Omega_{d - 1} = 0. \label{Sphere integration}
\end{equation}
And recalling the integrals $ \mathcal{Y}_{i, j} (\alpha, \beta) $
\begin{equation}
  \mathcal{Y}_{i, j} (\alpha, \beta) = \frac{\mathfrak{g}}{\lambda_C^d}
  \mathcal{A}_{d - 1} (m c^2)^{i + j - d} \int_0^{\infty} \frac{\sinh^i
  \vartheta \cosh^j \vartheta}{\re^{\alpha + \beta m c^2 \cosh \vartheta} -
  \varsigma} \mathrm{d} \vartheta .
\end{equation}
Using Eqs. \eqref{p parameter}-\eqref{Sphere integration}, we can further calculate
\begin{align}
  c \int \ensuremath{\boldsymbol{\varpi}} p^{\mu} f^{(0)}  = &~ \mathcal{Y}_{d
  - 1, 1} U^{\mu}, \\
  c \int \ensuremath{\boldsymbol{\varpi}} p^{\mu} p^{\sigma} f^{(0)}  = &~
  \mathcal{Y}_{d - 1, 2}  \frac{1}{c^2} U^{\mu} U^{\sigma} + \frac{1}{d}
  \mathcal{Y}_{d + 1, 0} \Delta^{\mu \sigma},
\end{align}
\begin{align}
  c \int \ensuremath{\boldsymbol{\varpi}}  \frac{1}{\varepsilon} p^{\mu}
  p^{\sigma} f^{(0)}  = &~ \mathcal{Y}_{d - 1, 1}  \frac{1}{c^2} U^{\mu}
  U^{\sigma} + \frac{1}{d} \mathcal{Y}_{d + 1, - 1} \Delta^{\mu \sigma}, \\
  c \int \ensuremath{\boldsymbol{\varpi}}  \frac{1}{\varepsilon} p^{\mu}
  p^{\nu} p^{\sigma} f^{(0)}  = &~ \mathcal{Y}_{d - 1, 2}  \frac{1}{c^4}
  U^{\mu} U^{\nu} U^{\sigma} + \frac{3}{d} \mathcal{Y}_{d + 1, 0}
  \frac{1}{c^2} U^{(\mu} \Delta^{\nu \sigma)}, \\
  c \int \ensuremath{\boldsymbol{\varpi}}  \frac{1}{\varepsilon} p^{\mu}
  p^{\nu} p^{\sigma} p^{\rho} f^{(0)}  = &~ \mathcal{Y}_{d - 1, 3}
  \frac{1}{c^6} U^{\mu} U^{\nu} U^{\sigma} U^{\rho} + \frac{6}{d}
  \mathcal{Y}_{d + 1, 1}  \frac{1}{c^4} U^{(\mu} U^{\nu} \Delta^{\sigma \rho)}
  \nonumber\\
    & + \frac{3}{(d + 2) d} \mathcal{Y}_{d + 3, - 1}  \frac{1}{c^2}
  \Delta^{(\mu \nu} \Delta^{\sigma \rho)},
\end{align}
\begin{align}
  c \int \ensuremath{\boldsymbol{\varpi}} \frac{1}{\varepsilon^2} p^{\mu}
  p^{\sigma} f^{(0)}  = &~ \mathcal{Y}_{d - 1, 0}  \frac{1}{c^2} U^{\mu}
  U^{\sigma} + \frac{1}{d} \mathcal{Y}_{d + 1, - 2} \Delta^{\mu \sigma}, \\
  c \int \ensuremath{\boldsymbol{\varpi}}  \frac{1}{\varepsilon^2} p^{\mu}
  p^{\nu} p^{\sigma} f^{(0)}  = &~ \mathcal{Y}_{d - 1, 1}  \frac{1}{c^4}
  U^{\mu} U^{\nu} U^{\sigma} + \frac{3}{d} \mathcal{Y}_{d + 1, - 1}
  \frac{1}{c^2} U^{(\mu} \Delta^{\nu \sigma)}, \\
  c \int \ensuremath{\boldsymbol{\varpi}}  \frac{1}{\varepsilon^2} p^{\mu}
  p^{\nu} p^{\sigma} p^{\rho} f^{(0)}  = &~ \mathcal{Y}_{d - 1, 2}
  \frac{1}{c^6} U^{\mu} U^{\nu} U^{\sigma} U^{\rho} + \frac{6}{d}
  \mathcal{Y}_{d + 1, 0}  \frac{1}{c^4} U^{(\mu} U^{\nu} \Delta^{\sigma \rho)}
  \nonumber\\
    & + \frac{3}{(d + 2) d} \mathcal{Y}_{d + 3, - 2}  \frac{1}{c^2}
  \Delta^{(\mu \nu} \Delta^{\sigma \rho)}, \\
  c \int \ensuremath{\boldsymbol{\varpi}}  \frac{1}{\varepsilon^2} p^{\mu}
  p^{\nu} p^{\sigma} p^{\rho} p^{\lambda} f^{(0)}  = &~ \mathcal{Y}_{d - 1, 3}
  \frac{1}{c^8} U^{\mu} U^{\nu} U^{\sigma} U^{\rho} U^{\lambda} +
  \frac{10}{d} \mathcal{Y}_{d + 1, 1}  \frac{1}{c^6} U^{(\mu} U^{\nu}
  U^{\sigma} \Delta^{\rho \lambda)} \nonumber\\
    & + \frac{15}{(d + 2) d} \mathcal{Y}_{d + 3, - 1}  \frac{1}{c^4}
  U^{(\mu} \Delta^{\nu \sigma} \Delta^{\rho \lambda)},
\end{align}
\begin{align}
  c \int \ensuremath{\boldsymbol{\varpi}}  \frac{1}{\varepsilon^3} p^{\mu}
  p^{\nu} p^{\sigma} f^{(0)}  = &~ \mathcal{Y}_{d - 1, 0}  \frac{1}{c^4}
  U^{\mu} U^{\nu} U^{\sigma} + \frac{3}{d} \mathcal{Y}_{d + 1, - 2}
  \frac{1}{c^2} U^{(\mu} \Delta^{\nu \sigma)}, \\
  c \int \ensuremath{\boldsymbol{\varpi}}  \frac{1}{\varepsilon^3} p^{\mu}
  p^{\nu} p^{\sigma} p^{\rho} f^{(0)}  = &~ \mathcal{Y}_{d - 1, 1}
  \frac{1}{c^6} U^{\mu} U^{\nu} U^{\sigma} U^{\rho} + \frac{6}{d}
  \mathcal{Y}_{d + 1, - 1}  \frac{1}{c^4} U^{(\mu} U^{\nu} \Delta^{\sigma
  \rho)} \nonumber\\
    & + \frac{3}{(d + 2) d} \mathcal{Y}_{d + 3, - 3}  \frac{1}{c^2}
  \Delta^{(\mu \nu} \Delta^{\sigma \rho)}, \\
  c \int \ensuremath{\boldsymbol{\varpi}}  \frac{1}{\varepsilon^3} p^{\mu}
  p^{\nu} p^{\sigma} p^{\rho} p^{\lambda} f^{(0)}  = &~ \mathcal{Y}_{d - 1, 2}
  \frac{1}{c^8} U^{\mu} U^{\nu} U^{\sigma} U^{\rho} U^{\lambda} +
  \frac{10}{d} \mathcal{Y}_{d + 1, 0}  \frac{1}{c^6} U^{(\mu} U^{\nu}
  U^{\sigma} \Delta^{\rho \lambda)} \nonumber\\
    & + \frac{15}{(d + 2) d} \mathcal{Y}_{d + 3, - 2}  \frac{1}{c^4}
  U^{(\mu} \Delta^{\nu \sigma} \Delta^{\rho \lambda)} .
\end{align}

Using the above integral formula and substituting Eqs.~\eqref{local equilibrium},
\eqref{f(1)}, and \eqref{f(2)} into the definitions of the particle flow vector
and the energy-momentum tensor \eqref{N,T}, we have
\begin{align}
  N^{\mu(0)} = & ~c \int \bm{\varpi} p^{\mu} f^{(0)}
  = ~ \mathcal{Y}_{d - 1, 1} U^{\mu}, \\
    &~  \nonumber\\
  T^{\mu \nu(0)} = &~ c \int \bm{\varpi} p^{\mu}
  p^{\nu} f^{(0)}
   = ~ \mathcal{Y}_{d - 1, 2}  \frac{1}{c^2} U^{\mu} U^{\nu} + \frac{1}{d}
  \mathcal{Y}_{d + 1, 0} \Delta^{\mu \nu},
\end{align}
\begin{align}
  N^{\mu(1)}  = &~ c \int \bm{\varpi} p^{\mu} f^{(1)}
  \nonumber\\
   = & - c^2 \tau \bigg[\Psi_{\sigma} \frac{\partial}{\partial \alpha} c \int
  \bm{\varpi}  \frac{1}{\varepsilon} p^{\mu} p^{\sigma}
  f^{(0)} - \Psi_{\rho \sigma} \frac{\partial}{\partial \alpha} c \int
  \bm{\varpi}  \frac{1}{\varepsilon} p^{\mu} p^{\rho}
  p^{\sigma} f^{(0)}\bigg] \nonumber\\
   = & - c^2 \tau \bigg[\frac{\partial}{\partial \alpha} \mathcal{Y}_{d - 1, 1}
  \frac{1}{c^2} U^{\mu} U^{\nu} \Psi_{\nu} + \frac{1}{d}
  \frac{\partial}{\partial \alpha} \mathcal{Y}_{d + 1, - 1} \Delta^{\mu \nu}
  \Psi_{\nu} \nonumber\\
    & - \frac{\partial}{\partial \alpha} \mathcal{Y}_{d - 1, 2}
  \frac{1}{c^4} U^{\mu} U^{\nu} U^{\sigma} \Psi_{\nu \sigma} - \frac{3}{d}
  \frac{\partial}{\partial \alpha} \mathcal{Y}_{d + 1, 0}  \frac{1}{c^2}
  U^{(\mu} \Delta^{\nu \sigma)} \Psi_{\nu \sigma}\bigg] \nonumber\\
   = & - c^2 \tau \bigg[\frac{\partial}{\partial \alpha} \mathcal{Y}_{d - 1, 1}
  \frac{1}{c^2} U^{\mu}  \dot{\alpha} + \frac{\partial}{\partial \alpha}
  \mathcal{Y}_{d - 1, 2}  \frac{1}{c^2} U^{\mu}  \dot{\beta} + \frac{1}{d}
  \frac{\partial}{\partial \alpha} \mathcal{Y}_{d + 1, 0}  \frac{1}{c^2}
  U^{\mu} \psi \nonumber\\
    & + \frac{1}{d}  \frac{\partial}{\partial \alpha} \mathcal{Y}_{d + 1, -
  1} \Delta^{\mu \nu} \mathcal{D}_{\nu} \alpha + \frac{1}{d}
  \frac{\partial}{\partial \alpha} \mathcal{Y}_{d + 1, 0}  \frac{1}{c^2} c^2
  \Delta^{\sigma \mu} \mathcal{D}_{\sigma} \beta \bigg],
\end{align}
\begin{align}
  T^{\mu \nu(1)}  = &~ c \int \bm{\varpi} p^{\mu}
  p^{\nu} f^{(1)} \nonumber\\
   = & - c^2 \tau \bigg[\Psi_{\sigma} \frac{\partial}{\partial \alpha} c \int
  \bm{\varpi}  \frac{1}{\varepsilon} p^{\mu} p^{\nu}
  p^{\sigma} f^{(0)} - \Psi_{\rho \sigma} \frac{\partial}{\partial \alpha} c
  \int \bm{\varpi}  \frac{1}{\varepsilon} p^{\mu} p^{\nu}
  p^{\rho} p^{\sigma} f^{(0)}\bigg] \nonumber\\
   = & - c^2 \tau \bigg[\frac{\partial}{\partial \alpha} \mathcal{Y}_{d - 1, 2}
  \frac{1}{c^4} U^{\mu} U^{\nu}  \dot{\alpha} + \frac{1}{d}
  \frac{\partial}{\partial \alpha} \mathcal{Y}_{d + 1, 0}  \frac{1}{c^2}
  (\Delta^{\mu \nu} \dot{\alpha} + 2 U^{(\mu} \Delta^{\nu) \sigma}
  \mathcal{D}_{\sigma} \alpha) \nonumber\\
    & + \frac{\partial}{\partial \alpha} \mathcal{Y}_{d - 1, 3}
  \frac{1}{c^4} U^{\mu} U^{\nu}  \dot{\beta} \nonumber\\
    & + \frac{1}{d}  \frac{\partial}{\partial \alpha} \mathcal{Y}_{d + 1, 1}
  \frac{1}{c^4}  (U^{\mu} U^{\nu} \psi + c^2 \Delta^{\mu \nu} \dot{\beta} + 2
  c^2 U^{(\mu} \Delta^{\nu) \rho} \mathcal{D}_{\rho} \beta) \nonumber\\
    & + \frac{1}{(d + 2) d}  \frac{\partial}{\partial \alpha} \mathcal{Y}_{d
  + 3, - 1}  \frac{1}{c^2} \left( \Delta^{\mu \nu} \psi + 2 \left( \Delta^{\mu
  (\rho} \Delta^{\sigma) \nu} \psi_{\rho \sigma} + \frac{1}{d} \psi
  \Delta^{\mu \nu} \right) \right)\bigg] \nonumber\\
   = & - c^2 \tau \bigg[\frac{\partial}{\partial \alpha} \mathcal{Y}_{d - 1, 2}
  \frac{1}{c^4} U^{\mu} U^{\nu}  \dot{\alpha} + \frac{\partial}{\partial
  \alpha} \mathcal{Y}_{d - 1, 3}  \frac{1}{c^4} U^{\mu} U^{\nu}  \dot{\beta} +
  \frac{1}{d}  \frac{\partial}{\partial \alpha} \mathcal{Y}_{d + 1, 1}
  \frac{1}{c^4} U^{\mu} U^{\nu} \psi \nonumber\\
    & + \frac{1}{d}  \frac{\partial}{\partial \alpha} \mathcal{Y}_{d + 1, 0}
  \frac{1}{c^2} 2 U^{(\mu} \Delta^{\nu) \sigma} \mathcal{D}_{\sigma} \alpha +
  \frac{1}{d}  \frac{\partial}{\partial \alpha} \mathcal{Y}_{d + 1, 1}
  \frac{1}{c^4} 2 c^2 U^{(\mu} \Delta^{\nu) \rho} \mathcal{D}_{\rho} \beta
  \nonumber\\
    & + \frac{1}{d}  \frac{\partial}{\partial \alpha} \mathcal{Y}_{d + 1, 0}
  \frac{1}{c^2} \Delta^{\mu \nu} \dot{\alpha} + \frac{1}{d}
  \frac{\partial}{\partial \alpha} \mathcal{Y}_{d + 1, 1}  \frac{1}{c^4} c^2
  \Delta^{\mu \nu} \dot{\beta} + \frac{1}{d^2}  \frac{\partial}{\partial
  \alpha} \mathcal{Y}_{d + 3, - 1}  \frac{1}{c^2} \Delta^{\mu \nu} \psi
  \nonumber\\
    & + \frac{2}{(d + 2) d}  \frac{\partial}{\partial \alpha} \mathcal{Y}_{d
  + 3, - 1}  \frac{1}{c^2} \Delta^{\mu (\rho} \Delta^{\sigma) \nu} \psi_{\rho
  \sigma}\bigg] ,
\end{align}
\begin{align}
  N^{\mu(2)}_{\text{EM}}  = &~ c \int \ensuremath{\boldsymbol{\varpi}}
  p^{\mu} f^{(2)}_{\text{EM}}\nonumber\\
   = &~ c^4 \tau^2  \bigg[q F^{\lambda}_{\enspace \rho} \Psi_{\lambda}
  \frac{\partial}{\partial \alpha} c \int \ensuremath{\boldsymbol{\varpi}}
  p^{\mu}  \frac{1}{\varepsilon^2} p^{\rho} f^{(0)} - 2 q
  F^{\lambda}_{\enspace \rho} \Psi_{\lambda \sigma} \frac{\partial}{\partial
  \alpha} c \int \ensuremath{\boldsymbol{\varpi}} p^{\mu}
  \frac{1}{\varepsilon^2} p^{\rho} p^{\sigma} f^{(0)}\nonumber\\
    & - q E_{\rho} \Psi_{\sigma}  \frac{\partial}{\partial \alpha} c \int
  \ensuremath{\boldsymbol{\varpi}} p^{\mu}  \frac{1}{\varepsilon^3} p^{\rho}
  p^{\sigma} f^{(0)} + q E_{\lambda} \Psi_{\rho \sigma}
  \frac{\partial}{\partial \alpha} c \int \ensuremath{\boldsymbol{\varpi}}
  p^{\mu}  \frac{1}{\varepsilon^3} p^{\lambda} p^{\rho} p^{\sigma} f^{(0)}\bigg]\nonumber\\
   = &~ c^4 \tau^2  \bigg[q F^{\lambda}_{\enspace \sigma} \Psi_{\lambda}
  \frac{\partial}{\partial \alpha} \mathcal{Y}_{d - 1, 0}  \frac{1}{c^2}
  U^{\mu} U^{\sigma} + \frac{1}{d} q F^{\lambda}_{\enspace \sigma}
  \Psi_{\lambda} \frac{\partial}{\partial \alpha} \mathcal{Y}_{d + 1, - 2}
  \Delta^{\mu \sigma}\nonumber\\
    & - 2 q F^{\lambda}_{\enspace \nu} \Psi_{\lambda \sigma}
  \frac{\partial}{\partial \alpha} \mathcal{Y}_{d - 1, 1}  \frac{1}{c^4}
  U^{\mu} U^{\nu} U^{\sigma} - \frac{3}{d} 2 q F^{\lambda}_{\enspace \nu}
  \Psi_{\lambda \sigma} \frac{\partial}{\partial \alpha} \mathcal{Y}_{d + 1, -
  1}  \frac{1}{c^2} U^{(\mu} \Delta^{\nu \sigma)}\nonumber\\
    & - q E_{\nu} \Psi_{\sigma}  \frac{\partial}{\partial \alpha}
  \mathcal{Y}_{d - 1, 0}  \frac{1}{c^4} U^{\mu} U^{\nu} U^{\sigma} -
  \frac{3}{d} q E_{\nu} \Psi_{\sigma}  \frac{\partial}{\partial \alpha}
  \mathcal{Y}_{d + 1, - 2}  \frac{1}{c^2} U^{(\mu} \Delta^{\nu \sigma)}\nonumber\\
    & + q E_{\nu} \Psi_{\rho \sigma}  \frac{\partial}{\partial \alpha}
  \mathcal{Y}_{d - 1, 1}  \frac{1}{c^6} U^{\mu} U^{\nu} U^{\sigma} U^{\rho}\nonumber\\
    & + \frac{6}{d} q E_{\nu} \Psi_{\rho \sigma}  \frac{\partial}{\partial
  \alpha} \mathcal{Y}_{d + 1, - 1}  \frac{1}{c^4} U^{(\mu} U^{\nu}
  \Delta^{\sigma \rho)} + \frac{3}{(d + 2) d} q E_{\nu} \Psi_{\rho \sigma}
  \frac{\partial}{\partial \alpha} \mathcal{Y}_{d + 3, - 3}  \frac{1}{c^2}
  \Delta^{(\mu \nu} \Delta^{\sigma \rho)}\bigg]\nonumber\\
   = &~ c^2 \tau^2 q \bigg[E^{\lambda} \mathcal{D}_{\lambda} \alpha \left(
  \frac{\partial}{\partial \alpha} \mathcal{Y}_{d - 1, 0} - \frac{1}{d}
  \frac{\partial}{\partial \alpha} \mathcal{Y}_{d + 1, - 2} \right) U^{\mu} +
  E^{\lambda} \mathcal{D}_{\lambda} \beta \frac{\partial}{\partial \alpha}
  \mathcal{Y}_{d - 1, 1} U^{\mu}\nonumber\\
    & + \frac{1}{d} c^2 \frac{\partial}{\partial \alpha} \mathcal{Y}_{d + 1,
  - 2} B^{\sigma \mu} \mathcal{D}_{\sigma} \alpha + \frac{1}{d} c^2
  \frac{\partial}{\partial \alpha} \mathcal{Y}_{d + 1, - 1} B^{\sigma \mu}
  \mathcal{D}_{\sigma} \beta\nonumber\\
    & + \frac{1}{d}  \frac{\partial}{\partial \alpha} \mathcal{Y}_{d + 1, -
  1} E^{\mu}  \dot{\beta} + \frac{1}{d^2} \left( 2 \frac{\partial}{\partial
  \alpha} \mathcal{Y}_{d + 1, - 1} - \frac{\partial}{\partial \alpha}
  \mathcal{Y}_{d + 3, - 3} \right) E^{\mu} \psi\nonumber\\
    & + \left( \frac{2}{d}  \frac{\partial}{\partial \alpha} \mathcal{Y}_{d
  + 1, - 1} - \frac{2}{(d + 2) d}  \frac{\partial}{\partial \alpha}
  \mathcal{Y}_{d + 3, - 3} \right) E^{\sigma} \psi_{\rho \sigma} \Delta^{\mu
  \rho}\bigg] ,
\end{align}
\begin{align}
  T^{\mu \nu(2)}_{\text{EM}}  = &~ c \int \ensuremath{\boldsymbol{\varpi}}
  p^{\mu} p^{\nu} f^{(2)}_{\text{EM}}\nonumber\\
   = &~ c^4 \tau^2 q \bigg[F^{\lambda}_{\enspace \rho} \Psi_{\lambda}
  \frac{\partial}{\partial \alpha} c \int \ensuremath{\boldsymbol{\varpi}}
  p^{\mu} p^{\nu}  \frac{1}{\varepsilon^2} p^{\rho} f^{(0)}\nonumber\\
    & - 2 F^{\lambda}_{\enspace \rho} \Psi_{\lambda \sigma}
  \frac{\partial}{\partial \alpha} c \int \ensuremath{\boldsymbol{\varpi}}
  p^{\mu} p^{\nu}  \frac{1}{\varepsilon^2} p^{\rho} p^{\sigma} f^{(0)}\nonumber\\
    & - E_{\rho} \Psi_{\sigma} \frac{\partial}{\partial \alpha} c \int
  \ensuremath{\boldsymbol{\varpi}} p^{\mu} p^{\nu}  \frac{1}{\varepsilon^3}
  p^{\rho} p^{\sigma} f^{(0)}\nonumber\\
    & + E_{\lambda} \Psi_{\rho \sigma} \frac{\partial}{\partial \alpha} c
  \int \ensuremath{\boldsymbol{\varpi}} p^{\mu} p^{\nu}
  \frac{1}{\varepsilon^3} p^{\lambda} p^{\rho} p^{\sigma} f^{(0)}\bigg]\nonumber\\
   = &~ c^4 \tau^2 q \bigg[F^{\lambda}_{\enspace \sigma} \Psi_{\lambda}
  \frac{\partial}{\partial \alpha} \mathcal{Y}_{d - 1, 1}  \frac{1}{c^4}
  U^{\mu} U^{\nu} U^{\sigma} + \frac{3}{d} F^{\lambda}_{\enspace \sigma}
  \Psi_{\lambda} \frac{\partial}{\partial \alpha} \mathcal{Y}_{d + 1, - 1}
  \frac{1}{c^2} U^{(\mu} \Delta^{\nu \sigma)}\nonumber\\
    & - 2 F^{\lambda}_{\enspace \rho} \Psi_{\lambda \sigma}
    \bigg(\frac{\partial}{\partial \alpha} \mathcal{Y}_{d - 1, 2}  \frac{1}{c^6}
  U^{\mu} U^{\nu} U^{\sigma} U^{\rho}\nonumber\\
    & + \frac{6}{d}  \frac{\partial}{\partial \alpha} \mathcal{Y}_{d + 1, 0}
  \frac{1}{c^4} U^{(\mu} U^{\nu} \Delta^{\sigma \rho)} + \frac{3}{(d + 2) d}
  \frac{\partial}{\partial \alpha} \mathcal{Y}_{d + 3, - 2}  \frac{1}{c^2}
  \Delta^{(\mu \nu} \Delta^{\sigma \rho)}\bigg)\nonumber\\
    & - E_{\rho} \Psi_{\sigma}  \bigg(\frac{\partial}{\partial \alpha}
  \mathcal{Y}_{d - 1, 1}  \frac{1}{c^6} U^{\mu} U^{\nu} U^{\sigma} U^{\rho}\nonumber\\
    & + \frac{6}{d}  \frac{\partial}{\partial \alpha} \mathcal{Y}_{d + 1, -
  1}  \frac{1}{c^4} U^{(\mu} U^{\nu} \Delta^{\sigma \rho)} + \frac{3}{(d + 2)
  d}  \frac{\partial}{\partial \alpha} \mathcal{Y}_{d + 3, - 3}  \frac{1}{c^2}
  \Delta^{(\mu \nu} \Delta^{\sigma \rho)}\bigg)\nonumber\\
    & + E_{\lambda} \Psi_{\rho \sigma} \bigg(\frac{\partial}{\partial \alpha}
  \mathcal{Y}_{d - 1, 2}  \frac{1}{c^8} U^{\mu} U^{\nu} U^{\sigma} U^{\rho}
  U^{\lambda}\nonumber\\
    & + \frac{10}{d}  \frac{\partial}{\partial \alpha} \mathcal{Y}_{d + 1,
  0}  \frac{1}{c^6} U^{(\mu} U^{\nu} U^{\sigma} \Delta^{\rho \lambda)} +
  \frac{15}{(d + 2) d}  \frac{\partial}{\partial \alpha} \mathcal{Y}_{d + 3, -
  2}  \frac{1}{c^4} U^{(\mu} \Delta^{\nu \sigma} \Delta^{\rho \lambda)}\bigg)\bigg]\nonumber\\
   = & \tau^2 q \bigg[\left( \frac{\partial}{\partial \alpha} \mathcal{Y}_{d - 1,
  1} - \frac{1}{d}  \frac{\partial}{\partial \alpha} \mathcal{Y}_{d + 1, - 1}
  \right) U^{\mu} U^{\nu} E^{\rho} \mathcal{D}_{\rho} \alpha +
  \frac{\partial}{\partial \alpha} \mathcal{Y}_{d - 1, 2} U^{\mu} U^{\nu}
  E^{\rho} \mathcal{D}_{\rho} \beta\nonumber\\
    & - \frac{2}{d}  \frac{\partial}{\partial \alpha} \mathcal{Y}_{d + 1, -
  1} c^2 U^{(\mu} B^{\nu) \sigma} \mathcal{D}_{\sigma} \alpha - \frac{2}{d}
  \frac{\partial}{\partial \alpha} \mathcal{Y}_{d + 1, 0} c^2
  \mathcal{D}_{\sigma} \beta U^{(\mu} B^{\nu) \sigma}\nonumber\\
    & + \frac{2}{d}  \frac{\partial}{\partial \alpha} \mathcal{Y}_{d + 1, 0}
  U^{(\mu} E^{\nu)} \dot{\beta} + \left( \frac{4}{d^2}
  \frac{\partial}{\partial \alpha} \mathcal{Y}_{d + 1, 0} - \frac{2}{d^2}
  \frac{\partial}{\partial \alpha} \mathcal{Y}_{d + 3, - 2} \right) U^{(\mu}
  E^{\nu)} \psi\nonumber\\
    & + \left( \frac{1}{d}  \frac{\partial}{\partial \alpha} \mathcal{Y}_{d
  + 1, - 1} - \frac{1}{d^2}  \frac{\partial}{\partial \alpha} \mathcal{Y}_{d +
  3, - 3} \right) c^2 \Delta^{\mu \nu} E^{\sigma} \mathcal{D}_{\sigma} \alpha
  + \frac{1}{d}  \frac{\partial}{\partial \alpha} \mathcal{Y}_{d + 1, 0} c^2
  \Delta^{\mu \nu} E^{\rho} \mathcal{D}_{\rho} \beta\nonumber\\
    & + \left( \frac{4}{d}  \frac{\partial}{\partial \alpha} \mathcal{Y}_{d
  + 1, 0} - \frac{4}{(d + 2) d}  \frac{\partial}{\partial \alpha}
  \mathcal{Y}_{d + 3, - 2} \right) E^{\rho} U^{(\mu} \Delta^{\nu) \sigma}
  \psi_{\rho \sigma}\nonumber\\
    & - \frac{4}{(d + 2) d}  \frac{\partial}{\partial \alpha} \mathcal{Y}_{d
  + 3, - 2} c^2 \Delta^{\sigma (\mu} B^{\nu) \rho}  \psi_{\sigma \rho}\nonumber\\
    & - \frac{2}{(d + 2) d}  \frac{\partial}{\partial \alpha} \mathcal{Y}_{d
  + 3, - 3} c^2 \left( \Delta^{\sigma (\mu} \Delta^{\nu) \rho} - \frac{1}{d}
  \Delta^{\mu \nu} \Delta^{\sigma \rho} \right) E_{\rho} \mathcal{D}_{\sigma}
  \alpha\bigg] ,
\end{align}
\begin{align}
  N_{\text{G}}^{\mu(2)} = &~ c \int \ensuremath{\boldsymbol{\varpi}} p^{\mu}
  \text{linear} [f^{(2)}_{\text{G}}]\nonumber\\
   = &~ c^4 \tau^2 c \int \ensuremath{\boldsymbol{\varpi}} p^{\mu}
  \frac{1}{\varepsilon^2} \bigg[p^{\sigma} p^{\rho}  \text{linear} [\nabla_{\sigma}
  \Psi_{\rho}] - p^{\sigma} p^{\rho} p^{\lambda}  \text{linear}
  [\nabla_{\sigma} \Psi_{\rho \lambda}]\nonumber\\
    & + \frac{1}{\varepsilon} \left( p^{\sigma} p^{\lambda} p^{\rho}
  \frac{1}{c^2} E^{(G)}_{(\lambda} U_{\rho)} \Psi_{\sigma} - p^{\rho}
  p^{\sigma} p^{\lambda} p^{\eta} \frac{1}{c^2} E^{(G)}_{(\lambda} U_{\eta)}
  \Psi_{\rho \sigma} \right)\bigg] \frac{\partial f^{(0)}}{\partial \alpha}\nonumber\\
   = &~ c^4 \tau^2  \bigg[\text{linear} [\nabla_{\sigma} \Psi_{\rho}]
  \frac{\partial}{\partial \alpha} c \int \ensuremath{\boldsymbol{\varpi}}
  \frac{1}{\varepsilon^2} p^{\mu} p^{\sigma} p^{\rho} f^{(0)}\nonumber\\
    & - \text{linear} [\nabla_{\sigma} \Psi_{\rho \lambda}]
  \frac{\partial}{\partial \alpha} c \int \ensuremath{\boldsymbol{\varpi}}
  \frac{1}{\varepsilon^2} p^{\mu} p^{\sigma} p^{\rho} p^{\lambda} f^{(0)}\nonumber\\
    & + \frac{1}{c^2} E^{(G)}_{(\lambda} U_{\rho)} \Psi_{\sigma}
  \frac{\partial}{\partial \alpha} c \int \ensuremath{\boldsymbol{\varpi}}
  \frac{1}{\varepsilon^3} p^{\mu} p^{\sigma} p^{\lambda} p^{\rho} f^{(0)}\nonumber\\
    & - \frac{1}{c^2} E^{(G)}_{(\lambda} U_{\eta)} \Psi_{\rho \sigma}
  \frac{\partial}{\partial \alpha} c \int \ensuremath{\boldsymbol{\varpi}}
  \frac{1}{\varepsilon^3} p^{\mu} p^{\rho} p^{\sigma} p^{\lambda} p^{\eta}
  f^{(0)}\bigg]\nonumber\\
   = &~ c^4 \tau^2  \bigg[\text{linear} [\nabla_{\sigma} \Psi_{\lambda}] \left(
  \frac{\partial}{\partial \alpha} \mathcal{Y}_{d - 1, 1}  \frac{1}{c^4}
  U^{\mu} U^{\lambda} U^{\sigma} + \frac{3}{d}  \frac{\partial}{\partial
  \alpha} \mathcal{Y}_{d + 1, - 1}  \frac{1}{c^2} U^{(\mu} \Delta^{\lambda
  \sigma)} \right)\nonumber\\
    & - \text{linear} [\nabla_{\sigma} \Psi_{\rho \lambda}]
    \bigg(\frac{\partial}{\partial \alpha} \mathcal{Y}_{d - 1, 2}  \frac{1}{c^6}
  U^{\mu} U^{\lambda} U^{\sigma} U^{\rho}\nonumber\\
    & + \frac{6}{d}  \frac{\partial}{\partial \alpha} \mathcal{Y}_{d + 1, 0}
  \frac{1}{c^4} U^{(\mu} U^{\lambda} \Delta^{\sigma \rho)} + \frac{3}{(d + 2)
  d}  \frac{\partial}{\partial \alpha} \mathcal{Y}_{d + 3, - 2}  \frac{1}{c^2}
  \Delta^{(\mu \lambda} \Delta^{\sigma \rho)}\bigg)\nonumber\\
    & + \frac{1}{c^2} E^{(G)}_{(\nu} U_{\rho)} \Psi_{\sigma}
    \bigg(\frac{\partial}{\partial \alpha} \mathcal{Y}_{d - 1, 1}  \frac{1}{c^6}
  U^{\mu} U^{\nu} U^{\sigma} U^{\rho}\nonumber\\
    & + \frac{6}{d}  \frac{\partial}{\partial \alpha} \mathcal{Y}_{d + 1, -
  1}  \frac{1}{c^4} U^{(\mu} U^{\nu} \Delta^{\sigma \rho)} + \frac{3}{(d + 2)
  d}  \frac{\partial}{\partial \alpha} \mathcal{Y}_{d + 3, - 3}  \frac{1}{c^2}
  \Delta^{(\mu \nu} \Delta^{\sigma \rho)}\bigg)\nonumber\\
    & - \frac{1}{c^2} E^{(G)}_{(\lambda} U_{\nu)} \Psi_{\rho \sigma}
    \bigg(\frac{\partial}{\partial \alpha} \mathcal{Y}_{d - 1, 2}  \frac{1}{c^8}
  U^{\mu} U^{\nu} U^{\sigma} U^{\rho} U^{\lambda}\nonumber\\
    & + \frac{10}{d}  \frac{\partial}{\partial \alpha} \mathcal{Y}_{d + 1,
  0}  \frac{1}{c^6} U^{(\mu} U^{\nu} U^{\sigma} \Delta^{\rho \lambda)} +
  \frac{15}{(d + 2) d}  \frac{\partial}{\partial \alpha} \mathcal{Y}_{d + 3, -
  2}  \frac{1}{c^4} U^{(\mu} \Delta^{\nu \sigma} \Delta^{\rho \lambda)}\bigg)\bigg]\nonumber\\
   = &~ \tau^2  \bigg[\left( \frac{\partial}{\partial \alpha} \mathcal{Y}_{d - 1,
  1} - \frac{1}{d}  \frac{\partial}{\partial \alpha} \mathcal{Y}_{d + 1, - 1}
  \right) U^{\mu} E^{\sigma}_{(G)} \mathcal{D}_{\sigma} \alpha +
  \frac{\partial}{\partial \alpha} \mathcal{Y}_{d - 1, 2} U^{\mu}
  E^{\sigma}_{(G)} \mathcal{D}_{\sigma} \beta\nonumber\\
    & - c^2 \frac{1}{d}  \frac{\partial}{\partial \alpha} \mathcal{Y}_{d +
  1, - 1} B^{\mu \sigma}_{(G)} \mathcal{D}_{\sigma} \alpha - c^2 \frac{1}{d}
  \frac{\partial}{\partial \alpha} \mathcal{Y}_{d + 1, 0} B^{\mu \sigma}_{(G)}
  \mathcal{D}_{\sigma} \beta\nonumber\\
    & + \frac{1}{d}  \frac{\partial}{\partial \alpha} \mathcal{Y}_{d + 1, 0}
  E^{\mu}_{(G)} \dot{\beta} + \frac{1}{d^2} \left( 2 \frac{\partial}{\partial
  \alpha} \mathcal{Y}_{d + 1, 0} - \frac{\partial}{\partial \alpha}
  \mathcal{Y}_{d + 3, - 2} \right) \psi E^{\mu}_{(G)}\nonumber\\
    & + \left( \frac{2}{d}  \frac{\partial}{\partial \alpha} \mathcal{Y}_{d
  + 1, 0} - \frac{2}{(d + 2) d} \frac{\partial}{\partial \alpha}
  \mathcal{Y}_{d + 3, - 2} \right) \psi_{\rho \sigma} E^{\rho}_{(G)}
  \Delta^{\sigma \mu}\bigg],
\end{align}
\begin{align}
  T^{\mu \nu(2)}_{\text{G}}  = &~ c \int \ensuremath{\boldsymbol{\varpi}}
  p^{\mu} p^{\nu} \text{linear} [f^{(2)}_{\text{G}}]\nonumber\\
   = &~ c^4 \tau^2 c \int \ensuremath{\boldsymbol{\varpi}} p^{\mu} p^{\nu}
  \frac{1}{\varepsilon^2} \bigg[p^{\sigma} p^{\rho}  \text{linear} [\nabla_{\sigma}
  \Psi_{\rho}] - p^{\sigma} p^{\rho} p^{\lambda}
  \text{linear} [\nabla_{\sigma} \Psi_{\rho \lambda}]\nonumber\\
    & - p^{\sigma} p^{\lambda} \frac{1}{c^2} E^{(G)}_{\lambda} \Psi_{\sigma}
  + p^{\rho} p^{\sigma} p^{\lambda} \frac{1}{c^2} E^{(G)}_{\lambda} \Psi_{\rho
  \sigma}\bigg] \frac{\partial f^{(0)}}{\partial \alpha}\nonumber\\
   = &~ c^4 \tau^2  \bigg[\text{linear} [\nabla_{\sigma} \Psi_{\rho}]
  \frac{\partial}{\partial \alpha} c \int \ensuremath{\boldsymbol{\varpi}}
  p^{\mu} p^{\nu}  \frac{1}{\varepsilon^2} p^{\sigma} p^{\rho} f^{(0)}\nonumber\\
    & - \text{linear} [\nabla_{\sigma} \Psi_{\rho \lambda}]
  \frac{\partial}{\partial \alpha} c \int \ensuremath{\boldsymbol{\varpi}}
  p^{\mu} p^{\nu}  \frac{1}{\varepsilon^2} p^{\sigma} p^{\rho} p^{\lambda}
  f^{(0)}\nonumber\\
    & - \frac{1}{c^2} E^{(G)}_{\lambda} \Psi_{\sigma}
  \frac{\partial}{\partial \alpha} c \int \ensuremath{\boldsymbol{\varpi}}
  p^{\mu} p^{\nu}  \frac{1}{\varepsilon^2} p^{\sigma} p^{\lambda} f^{(0)}\nonumber\\
    & + \frac{1}{c^2} E^{(G)}_{\lambda} \Psi_{\rho \sigma}
  \frac{\partial}{\partial \alpha} c \int \ensuremath{\boldsymbol{\varpi}}
  p^{\mu} p^{\nu}  \frac{1}{\varepsilon^2} p^{\rho} p^{\sigma} p^{\lambda}
  f^{(0)}\bigg]\nonumber\\
   = &~ c^4 \tau^2  \bigg[\text{linear} [\nabla_{\sigma} \Psi_{\rho}]
   \bigg(\frac{\partial}{\partial \alpha} \mathcal{Y}_{d - 1, 2}  \frac{1}{c^6}
  U^{\mu} U^{\nu} U^{\sigma} U^{\rho}\nonumber\\
    & + \frac{6}{d}  \frac{\partial}{\partial \alpha} \mathcal{Y}_{d + 1, 0}
  \frac{1}{c^4} U^{(\mu} U^{\nu} \Delta^{\sigma \rho)} + \frac{3}{(d + 2) d}
  \frac{\partial}{\partial \alpha} \mathcal{Y}_{d + 3, - 2}  \frac{1}{c^2}
  \Delta^{(\mu \nu} \Delta^{\sigma \rho)}\bigg)\nonumber\\
    & - \text{linear} [\nabla_{\sigma} \Psi_{\rho \lambda}]
    \bigg(\frac{\partial}{\partial \alpha} \mathcal{Y}_{d - 1, 3}  \frac{1}{c^8}
  U^{\mu} U^{\nu} U^{\sigma} U^{\rho} U^{\lambda}\nonumber\\
    & + \frac{10}{d}  \frac{\partial}{\partial \alpha} \mathcal{Y}_{d + 1,
  1}  \frac{1}{c^6} U^{(\mu} U^{\nu} U^{\sigma} \Delta^{\rho \lambda)} +
  \frac{15}{(d + 2) d}  \frac{\partial}{\partial \alpha} \mathcal{Y}_{d + 3, -
  1}  \frac{1}{c^4} U^{(\mu} \Delta^{\nu \sigma} \Delta^{\rho \lambda)}\bigg)\nonumber\\
    & - \frac{1}{c^2} E^{(G)}_{\rho} \Psi_{\sigma} \bigg(\frac{\partial}{\partial
  \alpha} \mathcal{Y}_{d - 1, 2}  \frac{1}{c^6} U^{\mu} U^{\nu} U^{\sigma}
  U^{\rho}\nonumber\\
    & + \frac{6}{d}  \frac{\partial}{\partial \alpha} \mathcal{Y}_{d + 1, 0}
  \frac{1}{c^4} U^{(\mu} U^{\nu} \Delta^{\sigma \rho)} + \frac{3}{(d + 2) d}
  \frac{\partial}{\partial \alpha} \mathcal{Y}_{d + 3, - 2}  \frac{1}{c^2}
  \Delta^{(\mu \nu} \Delta^{\sigma \rho)}\bigg)\nonumber\\
    & + \frac{1}{c^2} E^{(G)}_{\lambda} \Psi_{\rho \sigma}
    \bigg(\frac{\partial}{\partial \alpha} \mathcal{Y}_{d - 1, 3}  \frac{1}{c^8}
  U^{\mu} U^{\nu} U^{\sigma} U^{\rho} U^{\lambda}\nonumber\\
    & + \frac{10}{d}  \frac{\partial}{\partial \alpha} \mathcal{Y}_{d + 1,
  1}  \frac{1}{c^6} U^{(\mu} U^{\nu} U^{\sigma} \Delta^{\rho \lambda)} +
  \frac{15}{(d + 2) d}  \frac{\partial}{\partial \alpha} \mathcal{Y}_{d + 3, -
  1}  \frac{1}{c^4} U^{(\mu} \Delta^{\nu \sigma} \Delta^{\rho \lambda)}\bigg)\bigg]\nonumber\\
   = &~ \tau^2  \bigg[\left( \frac{\partial}{\partial \alpha} \mathcal{Y}_{d - 1,
  2} - \frac{1}{d}  \frac{\partial}{\partial \alpha} \mathcal{Y}_{d + 1, 0}
  \right) \frac{1}{c^2} U^{\mu} U^{\nu} E^{\sigma}_{(G)} \mathcal{D}_{\sigma}
  \alpha + \frac{\partial}{\partial \alpha} \mathcal{Y}_{d - 1, 3}
  \frac{1}{c^2} U^{\mu} U^{\nu} E^{\sigma}_{(G)} \mathcal{D}_{\sigma} \beta\nonumber\\
    & - \frac{2}{d}  \frac{\partial}{\partial \alpha} \mathcal{Y}_{d + 1, 0}
  U^{(\nu} B_{(G)}^{\mu) \sigma} \mathcal{D}_{\sigma} \alpha - \frac{2}{d}
  \frac{\partial}{\partial \alpha} \mathcal{Y}_{d + 1, 1}
  U^{(\nu} B^{\mu) \sigma}_{(G)} \mathcal{D}_{\sigma} \beta\nonumber\\
    & + \frac{2}{d}  \frac{\partial}{\partial \alpha} \mathcal{Y}_{d + 1, 1}
  \frac{1}{c^2}  \dot{\beta} E^{(\mu}_{(G)} U^{\nu)} + \left( \frac{4}{d^2}
  \frac{\partial}{\partial \alpha} \mathcal{Y}_{d + 1, 1} - \frac{2}{d^2}
  \frac{\partial}{\partial \alpha} \mathcal{Y}_{d + 3, - 1} \right)
  \frac{1}{c^2} \psi E^{(\mu}_{(G)} U^{\nu)}\nonumber\\
    & + \left( \frac{4}{d}  \frac{\partial}{\partial \alpha} \mathcal{Y}_{d
  + 1, 1} - \frac{4}{(d + 2) d}  \frac{\partial}{\partial \alpha}
  \mathcal{Y}_{d + 3, - 1} \right)  \frac{1}{c^2} \psi_{\rho \sigma}
  E^{\rho}_{(G)} \Delta^{\sigma (\mu} U^{\nu)}\nonumber\\
    & + \left( \frac{1}{d}  \frac{\partial}{\partial \alpha} \mathcal{Y}_{d
  + 1, 0} - \frac{1}{d^2}  \frac{\partial}{\partial \alpha} \mathcal{Y}_{d +
  3, - 2} \right) E^{\sigma}_{(G)} \mathcal{D}_{\sigma} \alpha \Delta^{\mu
  \nu} + \frac{1}{d}  \frac{\partial}{\partial \alpha} \mathcal{Y}_{d + 1, 1}
  E^{\sigma}_{(G)} \mathcal{D}_{\sigma} \beta \Delta^{\mu \nu}\nonumber\\
    & - \frac{4}{(d + 2) d}  \frac{\partial}{\partial \alpha} \mathcal{Y}_{d
  + 3, - 1} \psi_{\rho \sigma} \Delta^{\sigma (\nu} B^{\mu) \rho}_{(G)}\nonumber\\
    & - \frac{2}{(d + 2) d}  \frac{\partial}{\partial \alpha} \mathcal{Y}_{d
  + 3, - 2} \left( \Delta^{\sigma (\mu} \Delta^{\nu) \rho} - \frac{1}{d}
  \Delta^{\sigma \rho} \Delta^{\mu \nu} \right) E_{\rho}^{(G)}
  \mathcal{D}_{\sigma} \alpha\bigg],
\end{align}
where we have already substituted the linear response part \eqref{linear part1}
and \eqref{linear part2} in $f^{(2)}_{\text{G}}$.
Extracting the scalar, vectorial, and tensorial dissipation terms from the above result
and using the relation \eqref{partail_y relation}, Eqs.~\eqref{S-S}-\eqref{T-V} follow.

\end{document}